\documentclass[12pt, draftclsnofoot, onecolumn, romanappendices]{IEEEtran}

\usepackage{graphicx}
\usepackage{verbatim}
\usepackage{float}

\usepackage{times} 
\usepackage{amsmath}
\usepackage{amsfonts}
\usepackage{amssymb}
\usepackage{cite}
\usepackage{subfigure}
\usepackage{url}
\usepackage{balance}
\usepackage{enumerate}

\usepackage{color}
\usepackage{mathrsfs} 

\usepackage{hyperref}

\newtheorem{theorem}{Theorem}
\newtheorem{lemma}[theorem]{Lemma}
\newtheorem{proposition}[theorem]{Proposition}
\newtheorem{corollary}[theorem]{Corollary}
\newtheorem{remark}{Remark}
\def\X{\mathbf{X}}
\def\Y{\mathbf{Y}}
\def\XL{\mathbf{X}^L}
\def\YL{\mathbf{Y}^L}

\def\X_l{\mathbf{X}_\ell} 
\def\Y_l{\mathbf{Y}_\ell} 
\def\Xk{\mathbf{X}_\ell}
\def\Yk{\mathbf{Y}_\ell}
\def\xk{\mathbf{x}_\ell}
\def\yk{\mathbf{y}_\ell}

\def\T{{T}}

\def\tiltRv{V_{\ell,\theta}}
\def\tiltDist{\vartheta_{\theta,\tau}}

\def\gamoneL{\Upsilon_{1,\ell}}
\def\gamtwoL{\Upsilon_{2,\ell}}

\def\rcus{RCU\textsubscript{s}}

\def\Qexp{\Psi}
\def\E0{E_{0,\rho}}

\newcommand{\Rcrb}[1]{R_{1/2}^\text{cr}\lro{#1}}
\newcommand{\Rcr}[1]{R_{s}^\text{cr}\lro{#1}}

\newcommand{\lefto}{\mathopen{}\left}
\newcommand{\lro}[1]{\lefto({#1}\right)}					
\newcommand{\lrbo}[1]{\lefto \lbrace {#1} \right \rbrace}	
\newcommand{\lrho}[1]{\lefto [ {#1} \right ]}				
\newcommand{\lrvo}[1]{\lefto | {#1} \right |}				
\allowdisplaybreaks
\begin{document}
\sloppy
\IEEEoverridecommandlockouts
\title{Saddlepoint Approximations for\\ Short-Packet Wireless Communications}

\author{Alejandro~Lancho,~\IEEEmembership{Member,~IEEE,},
        Johan~\"Ostman,~\IEEEmembership{Student Member,~IEEE,}
        Giuseppe~Durisi,~\IEEEmembership{Senior Member,~IEEE,}
        Tobias Koch,~\IEEEmembership{Senior Member,~IEEE,}
         and~Gonzalo~Vazquez-Vilar,~\IEEEmembership{Member,~IEEE}

	\thanks{A.~Lancho has received funding from the European Research Council (ERC) under the European Union's Horizon 2020 research and innovation programme (grant agreement number 714161) and from the Wallenberg AI, autonomous systems and software program. J.~\"Ostman and G. Durisi have been supported by the Swedish Research Council under grants 2014-6066 and 2016-03293. T.~Koch has received funding from the European Research Council (ERC) under the European Union's Horizon 2020 research and innovation programme (grant agreement number 714161) and from the Spanish Ministerio de Econom\'ia y Competitividad under grants RYC-2014-16332 and TEC2016-78434-C3-3-R (AEI/FEDER, EU). G.~Vazquez-Vilar has received funding from the European Research Council (ERC) under the European Union's Horizon 2020 research and innovation programme (grant agreement number 714161) and from the Spanish Ministerio de Econom\'ia y Competitividad under grant TEC2016-78434-C3-3-R (AEI/FEDER, EU).
	The material in this paper was presented in part at the IEEE International Symposium on Information Theory, Paris, France, July 2019, and at the 53rd Annual Asilomar Conference on Signals, Systems, and Computers, Pacific Grove, CA, USA, November 2019. 

	A.~Lancho, J.~\"Ostman, and G.~Durisi are with the Department of Electrical Engineering, Chalmers University of Technology, Gothenburg, Sweden (e-mails: lancho@ieee.org; johanos@chalmers.se; durisi@chalmers.se).
	
	T.~Koch and G. Vazquez-Vilar are with the Signal Theory and Communications Department, Universidad Carlos~III de Madrid, Spain, and with the Gregorio Mara\~n\'on Health Research Institute, Madrid, Spain (e-mails: koch@tsc.uc3m.es; gonzalo.vazquez@uc3m.es).}
	
}

\maketitle

\begin{abstract}
In recent years, the derivation of nonasymptotic converse and achievability bounds on the maximum coding rate as a function of the error probability and blocklength has gained attention in the information theory literature. While these bounds are accurate for many scenarios of interest, they need to be evaluated numerically for most wireless channels of practical interest, and their evaluation is computationally demanding. This paper presents saddlepoint approximations of state-of-the-art converse and achievability bounds for noncoherent, single-antenna, Rayleigh block-fading channels. These approximations can be calculated efficiently and are shown to be accurate for SNR values as small as 0~dB and blocklengths of 168 channel uses or more. 
\end{abstract}
\section{Introduction}
One of the main goals in the design of communication systems is to maximize the transmission rate (also referred to as coding rate) under given constraints on system parameters, such as power, bandwidth, latency, or reliability. In the absence of stringent latency constraints, which is the case for traditional wireless communication systems, the transmission rate can be improved by increasing the length of the packets sent to convey information from a transmitter to a receiver. We shall refer to the packet length as \emph{blocklength}. This implies that asymptotic information-theoretical metrics, such as \emph{capacity} or \emph{outage capacity}, are good benchmarks. Indeed, capacity is defined as the maximum coding rate at which data can be transmitted with vanishing error probability when the blocklength tends to infinity. Similarly, outage capacity is defined as the maximum coding rate at which data can be transmitted for a given error probability when the blocklength tends to infinity. These metrics have been widely used as performance benchmarks in the design of coding schemes for traditional wireless communication systems.

Next generation wireless communication systems will support services and applications requiring \textit{ultra-reliable low-latency communications} (URLLC) \cite{Durisi_Koch_Popovski_2015,Popovski_2018}.
In URLLC, the devices exchange packets at low rates aiming for probabilities of error less than or equal to $10^{-5}$ and satisfying stringent latency constraints \cite{Popovski_2018}. This is typically achieved by transmitting short packets. Recalling that capacity and outage capacity are defined under the assumption that the blocklength tends to infinity, it follows that for latency-constrained communications a more refined characterization of the maximum coding rate as a function of the blocklength is needed.
\subsection{State of the Art}\label{sec:literature}
For most channel models of interest, obtaining a closed-form expression of the maximum coding rate is out of reach. Hence, there are two main directions to characterize the maximum coding rate as a function of the blocklength:

\begin{enumerate}[i)]
\item \textbf{Nonasymptotic bounds}: By obtaining upper and lower bounds on the maximum coding rate,
 the  area in which this rate lies for a specific error probability and blocklength can be characterized. Often such bounds are expressed in terms of tail probabilities of sums of independent and identically distributed (i.i.d.) random variables and need to be evaluated numerically through computationally demanding procedures. Nonasymptotic bounds for several channel models can be found, e.g., in \cite{Polyanskiy_Poor_Verdu,Short_Packets_Durisi_Koch_TCOM_2015}. The nonasymptotic bounds considered in this paper are the meta-converse (MC) bound~\cite[Th. 31]{Polyanskiy_Poor_Verdu}  and the random coding union bound with parameter $s$ (\rcus{})~\cite[Th. 1]{Martinez_Guillen_i_Fabregas_2011}. The MC bound and the \rcus{} bound yield an upper and a lower bound on the maximum coding rate for a fixed error probability and blocklength, respectively.
\item \textbf{Refined asymptotic expansions}:\footnote{We refer to these expansion as \emph{refined} asymptotic expansions to better differentiate them from capacity or outage capacity, which are first-order asymptotic expansions of the maximum coding rate.} Perform asymptotic expansions of the maximum coding rate or the error probability that become accurate as the blocklength grows. Typically, such expansions are available in closed form and describe how the maximum coding rate converges to capacity or how the error probability vanishes as the blocklength increases for a fixed coding rate.
\end{enumerate}

The refined asymptotic expansions are usually obtained by expanding the tail probabilities appearing in the aforementioned nonasymptotic bounds. One possibility is to fix a reliability constraint and study the maximum coding rate as a function of the blocklength in the limit as the blocklength tends to infinity.
This approach was followed \emph{inter alia} by Polyanskiy \emph{et al.}~\cite{Polyanskiy_Poor_Verdu} who showed, for various channels with positive finite capacity $C$, that the maximum coding rate $R^*(n,\epsilon)$ at which data can be transmitted using an error-correcting code of a fixed blocklength~$n$, with a block-error probability not larger than $\epsilon$, can be expanded as
\begin{equation}\label{eq:rate_ppv}
R^*(n,\epsilon)=C-\sqrt{\frac{V}{n}}Q^{-1}(\epsilon)+\mathcal{O}\left(\frac{\log n}{n}\right)
\end{equation}
where $V$ denotes the channel dispersion, $Q^{-1}(\cdot)$ denotes the inverse of the Gaussian $Q$-function, and $\mathcal{O}\left((\log n)/n\right)$ comprises terms that decay no slower than $(\log n)/n$. The approximation that follows from \eqref{eq:rate_ppv} by omitting the $\mathcal{O}((\log n)/n)$ term is sometimes referred to as \emph{normal approximation}. The normal approximation has been established as a benchmark for short error-correcting codes. For example, \cite{Mustafa2019} compares the performance of  codes of blocklength $n=128$ against the normal approximation of a Gaussian channel. The normal approximation further serves as a proxy for the maximum coding rate in the analysis and optimization of communication systems that exchange short packets. For example, the normal approximation has appeared in numerous papers on short-packet wireless communications, including \cite{MakkiSvenssonZorzi2014,MakkiSvenssonZorzi2015,MakkiSvenssonZorzi2016,Haghifam2017,YuChenLiDingVucetic2018,SunYanYangDingShenZhong2018,ZhangLiang2018,WangDingPoor2019,LiYangYan2019}.

The work by Polyanskiy \emph{et al.} \cite{Polyanskiy_Poor_Verdu} has been generalized to several wireless communication channels; see, e.g., \cite{Polyanskiy_2011, Yang_2014,Hoydis_2015,Collins_Polyanskiy_2018,LKD17}. Particularly relevant for this work is the high-SNR normal approximation for noncoherent single-antenna Rayleigh block-fading channels derived in~\cite{LKD17}. By means of numerical examples, it was shown in~\cite{LKD17} that this normal approximation is accurate for probabilities of error above $10^{-3}$ and SNR values larger than or equal to $15$~dB.

A second option to obtain refined asymptotic expansions is to fix the coding rate and to study the exponential decay of the error probability as the blocklength grows to infinity. 
Error exponent results for block-fading channels can be found in, e.g., \cite{Telatar99,Wu2007,Shin2009,abou-faycal99-a,Ostman_2018}. 

In general, normal approximations are accurate for moderate error probabilities and when the rate is close to capacity. In contrast, error exponents are accurate for moderate coding rates when the probability of error is close to zero.  URLLC services operate at error probabilities of around $10^{-5}$ and SNR values of around $0$~dB \cite{Popovski_2018}. For these values, both normal approximations and error exponents may become inaccurate, in which case they constitute poor benchmarks for short error-correcting codes, and using them as a proxy for the maximum coding rate or the error probability in the analysis and optimization of short-packet communication systems may give rise to misleading results. Consequently, there is a need for refined approximations that characterize the coding rate for error probabilities below $10^{-5}$ and SNR values close to $0$~dB.  
\subsection{Contributions}\label{sec:contributions}
To provide approximations that are accurate in regimes where neither normal approximations nor error exponents are, this paper makes use of the classical \emph{saddlepoint method}~\cite[Ch.~XVI]{Feller},\cite{Jensen_saddlepoint}. The saddlepoint method has been applied in \cite{Scarlett_2014} to obtain approximations of  the \rcus{} bound and the random coding union (RCU) bound \cite[Th. 16]{Polyanskiy_Poor_Verdu} for channels with finite alphabets. It has been further applied in \cite{Font-Segura_CISS18, Vazquez-Vilar_ISIT18} to obtain approximations of the RCU bound and MC bound for symmetric memoryless channels. Similar approximations based on Laplace integration were used in \cite{Tomaso_2015,Tomaso_2016} to approximate the RCU bound and MC bound for Gaussian channels.

\emph{Saddlepoint expansions} of nonasymptotic bounds can be obtained by applying the saddlepoint method to expand the tail probabilities of sums of i.i.d. random variables appearing in these bounds. These random variables typically depend on system parameters such as the SNR.  Since existing saddlepoint methods ignore such dependencies, the error terms in the saddlepoint expansions also depend on these parameters and may even be unbounded in them. This is particularly problematic if the saddlepoint expansions are the starting point for \mbox{asymptotic analyses}.

To overcome this problem, we derive in Section~\ref{sec:saddle_exp} saddlepoint expansions for random variables whose distribution depends on a system parameter $\theta$, carefully analyze the error terms, and demonstrate that they are uniform in $\theta$.
The obtained expansions are then applied to the tail probabilities appearing in the nonasymptotic MC and \rcus{} bounds, leading to saddlepoint expansions of the MC and \rcus{} bounds with error terms that depend only on the blocklength and are uniform in the remaining parameters. By means of numerical analyses, we show that the \emph{saddlepoint approximations} that follow by ignoring the error terms in the saddlepoint expansions are accurate, i.e., they are indistinguishable from the corresponding nonasymptotic bounds, if the SNR is greater than or equal to $0$~dB and if the probability of error is larger than or equal to $10^{-8}$. 
Furthermore, they require a much lower computational cost than the corresponding nonasymptotic bounds.
Finally, we show that the normal approximation and the error exponent can be recovered from the saddlepoint expansions. They thus provide a unifying characterization of the two regimes, which are usually considered separately in the literature.

The rest of this paper is organized as follows. In Section~\ref{sec:system}, we present the system model. In Section~\ref{sec:nonasym}, we introduce the most important quantities and the nonasymptotic bounds used in the paper. In Section~\ref{sec:saddle_exp}, we derive saddlepoint expansions for random variables whose distribution depends on a parameter $\theta$. In Section~\ref{sec:saddle_RBF}, we apply those saddlepoint expansions to the nonasymptotic bounds presented in Section~\ref{sec:nonasym}. In Section~\ref{sec:NA_err}, we present normal approximations and error-exponent approximations that can be obtained from the saddlepoint expansions. In Section~\ref{sec:num_results}, we present numerical examples that help to assess the accuracy of the presented approximations. In Section~\ref{sec:discussion}, we discuss the computational complexity and accuracy of the proposed saddlepoint approximations, and we compare them with the nonasymptotic bounds as well as the refined asymptotic expansions available in the literature. Section~\ref{sec:conclusions} concludes the paper. Some of the proofs are deferred to the appendices.
\section{System Model}
\label{sec:system}
\begin{figure}[t!]
  \centering
\includegraphics[width=0.9\columnwidth]{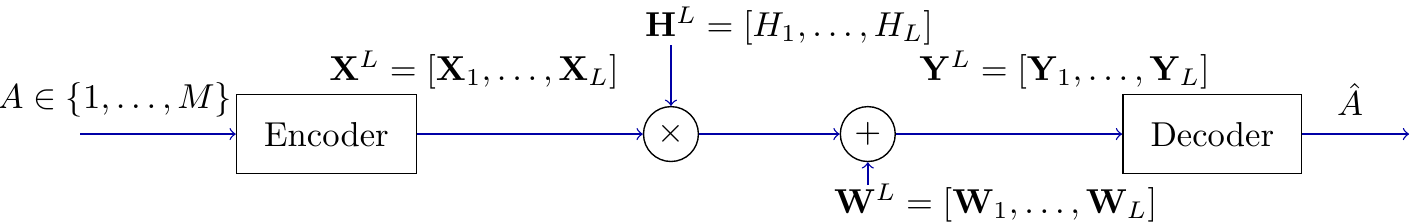}
\vspace{-1em}
\caption{Schema of the communication system with the single-antenna Rayleigh block-fading channel as channel model.}
\label{fig:system}
\end{figure}
The system and channel model is illustrated in Fig.~\ref{fig:system}. We consider a single-antenna Rayleigh block-fading channel with coherence interval $\T$, a channel model that is commonly used in the literature; see e.g., \cite{Marzetta_99,Zheng_2002}. For this channel model, the input-output relation within the $\ell$-th coherence interval is given by
\begin{equation}
\label{ch_model}
\Y_l = H_{\ell}\X_l + \mathbf{W}_{\ell}
\end{equation}
where $\X_l$ and $\Y_l$ are $\T$-dimensional, complex-valued, random vectors containing the input and output signals, respectively;
$\mathbf{W}_{\ell}$ is the additive noise with i.i.d., zero-mean, unit-variance, circularly-symmetric, complex Gaussian entries;
and $H_{\ell}$ is a zero-mean, unit-variance, circularly-symmetric, complex Gaussian random variable.
We assume that $H_{\ell}$ and $\mathbf{W}_{\ell}$ are independent and take on independent realizations over successive coherence intervals.
Moreover, the joint law of $(H_{\ell},\mathbf{W}_{\ell})$ does not depend on the channel inputs.
We consider a noncoherent setting where the transmitter and the receiver are aware of the distribution of $H_{\ell}$ but not of its realization.\footnote{The assumption that $H_{\ell}$ and $\mathbf{W}_{\ell}$ take on independent realizations over successive coherence intervals is critical for the results obtained in this paper, namely, Proposition~\ref{pr:main_saddle} in Section~\ref{sec:saddle_exp} and the saddlepoint expansions presented in Section~\ref{sec:saddle_RBF} that follow from it. In contrast, the assumption that $H_{\ell}$ is zero-mean Gaussian is not critical since Proposition~\ref{pr:main_saddle} applies to any fading distribution for which \eqref{eq:MGF_unif_bounded} and \eqref{eq:inf_second_positive} are satisfied.}

Note that, by considering a noncoherent setting, we do not preclude the possibility of estimating the fading coefficients, e.g., by transmitting pilot symbols. Instead, as in \cite{Lapidoth2005}, we view the transmission of pilot symbols as a special case of coding and the estimation of the fading coefficients as part of the decoder. By studying the maximum coding rate of the noncoherent Rayleigh block-fading channel, we thus characterize the largest transmission rate at which data can be transmitted, irrespective of the employed channel-estimation method. See also \cite{Durisi_Koch_Popovski_2015}.

We next introduce the notion of a channel code. For simplicity, we shall restrict ourselves to codes whose blocklength $n$ satisfies $n=L\T$, where $L$ denotes the number of coherence intervals of length $\T$ needed to transmit the entire codeword.
An $(M,L,T,\epsilon,\rho)$-code for the channel \eqref{ch_model} consists of:
\begin{enumerate}
	\item An encoder $f$: $\{1,\dots,M\}\rightarrow\mathbb{C}^{L\T}$ that maps a message $A$, which is uniformly distributed on $\{1,\dots,M\}$, to a codeword $\XL=\left[\mathbf{X}_1,\dots,\mathbf{X}_L\right]=f(A)$. The codewords satisfy the power constraint\footnote{While in the information and communication theory literature, it is more common to impose a power constraint per codeword $\mathbf{X}^L$, practical systems typically require a per-coherence-interval constraint. Although it may be preferable to impose \eqref{power_const} with inequality, since it allows more freedom in optimizing the codebook, it seems plausible that using maximum power is optimal. For the high-SNR normal approximation presented in \cite{LKD17}, this turns out to be the case.}
	\begin{equation}
	\label{power_const}
	\|\X_l\|^2 = \T\rho, \quad \ell=1,\dots,L.
	\end{equation}
	Since the variance of $H_{\ell}$ and of the entries of $\mathbf{W}_{\ell}$ are normalized to one, $\rho$ in \eqref{power_const} can be interpreted as the average SNR at the receiver.
	\item A decoder $g$: $\mathbb{C}^{L\T}\rightarrow\{1,\dots,M\}$ satisfying the average error probability constraint $\mathsf{P}\lrho{ g\lro{\YL}\neq A}\leq\epsilon$, where $\YL=\left[\mathbf{Y}_1,\dots,\mathbf{Y}_L\right]$ is the channel output induced by the transmitted codeword $\XL  = f(A)$ according to \eqref{ch_model}.
\end{enumerate}
The \emph{maximum coding rate} and the \emph{minimum error probability} are respectively defined as\footnote{Here and throughout this paper, $\log(\cdot)$ denotes the natural logarithm. Thus, rates have the dimension nats per channel use.}
\begin{subequations}
	\begin{IEEEeqnarray}{rCl}
		R^*\lro{L,T,\epsilon,\rho} &\triangleq& \sup\lrbo{\frac{\log M}{L\T}\,:\,\exists(M,L,T,\epsilon,\rho)\text{-code}}\IEEEeqnarraynumspace\\
		\epsilon^*\lro{L,T,R,\rho} &\triangleq& \inf\lrbo{\epsilon: \,\exists(e^{L\T R},L,T,\epsilon,\rho)\text{-code}}. \IEEEeqnarraynumspace
	\end{IEEEeqnarray}
\end{subequations}
In words, $R^*\lro{L,T,\epsilon,\rho}$ denotes the largest transmission rate at which data can be transmitted with an error probability not exceeding $\epsilon$ using a channel code of blocklength $LT$. Likewise, $\epsilon^*\lro{L,T,R,\rho}$ denotes the smallest error probability at which data can be transmitted at rate $R$ using a channel code of blocklength $LT$. In this paper, we shall present our results in terms of error probabilities and use that upper (lower) bounds on $\epsilon^*\lro{L,T,R,\rho}$ can be translated into lower (upper) bounds on $R^*\lro{L,T,\epsilon,\rho}$ and \emph{vice versa}.
\section{Preliminaries and Nonasymptotic Bounds}\label{sec:nonasym}
\subsection{Preliminaries}\label{sec:preliminaries}
According to \eqref{ch_model}, conditioned on $\XL=\mathbf{x}^L$, the output vector $\YL$ is blockwise i.i.d.\ Gaussian. Thus, the conditional probability density function (pdf) of $\Yk$ given $\Xk=\mathbf{x}$ is independent of~$\ell$ and satisfies
\begin{equation}
\label{cond_pdf}
\text{p}_{\mathbf{Y}|\mathbf{X}}(\mathbf{y}|\mathbf{x})={\frac{1}{\pi^\T(1+\|\mathbf{x}\|^2)}\exp\biggl\{-\|\mathbf{y}\|^2+\frac{|\mathbf{y}^H\mathbf{x}|^2}{1+\|\mathbf{x}\|^2}\biggr\}}.
\end{equation}
Here and throughout the paper, we omit the subscript $\ell$ when immaterial.

We shall evaluate the achievability bounds for inputs of the form $\XL=\sqrt{\T\rho}\mathbf{U}^L$, where the components of $\mathbf{U}^L=\left[\mathbf{U}_1,\dots,\mathbf{U}_L\right]$ are i.i.d. and uniformly distributed on the unit sphere in $\mathbb{C}^\T$. This distribution of $\XL$ can be viewed as a single-antenna particularization of \emph{unitary space-time modulation} (USTM) \cite{Hochwald_2000} and is capacity achieving for the power constraint \eqref{power_const}.\footnote{It is unclear whether USTM is also optimal at finite blocklength. However, for the blocklengths considered in the numerical examples of Section~\ref{sec:num_results}, the \rcus{} bound, computed for USTM inputs, is close to the MC bound, which applies to all input distributions satisfying \eqref{power_const}. This suggests that USTM is  at least close to optimal at finite blocklength.} In the following, we shall write $\bar{\text{P}}_{\mathbf{X}}$ to denote the distribution of $\X_l=\sqrt{\T\rho} \mathbf{U}_{\ell}$.

For the USTM input distribution, we define the \emph{generalized information density} as
\begin{IEEEeqnarray}{rCll}
	i_s(\xk;\xk)  &\triangleq&  \log \frac{\text{p}_{\Y_l|\X_l}(\yk|\xk)^s}{\int \text{p}_{\Y_l|\X_l}(\yk|\tilde{\mathbf{x}})^s \mathrm{d} \bar{\text{P}}_{\mathbf{X}}(\tilde{\mathbf{x}})}, \quad & s>0. \label{eq:inf_dens_summand_def}
\end{IEEEeqnarray}
Introducing the generalized information density is convenient because its cumulant generating function (CGF) is closely related to Gallager's $E_0(\rho,s)$ function, which plays an important role in the analysis of error exponents; see \cite{Martinez_Guillen_i_Fabregas_2011} for more details.
It can be shown that $i_s(\Xk;\Yk)$ depends on $\Xk$ only via $\|\Xk\|^2 = \T\rho$. Conditioned on $\|\Xk\|^2 = \T\rho$, we obtain that $i_s(\Xk;\Yk)$ can be written as 
\begin{multline}
	i_s(\Xk;\Yk)  \stackrel{\scriptstyle{\mathscr{L}}}{=}  \lro{\T-1}\log(s\T\rho)-\log\Gamma(\T)- s\frac{\T\rho \gamtwoL}{1+\T\rho}\\
{}+(\T-1)\log\lro{\frac{(1+\T\rho)\gamoneL+\gamtwoL}{1+\T\rho}}-\log\tilde{\gamma}\lro{\T-1,s\frac{\T\rho((1+\T\rho)\gamoneL+\gamtwoL)}{1+\T\rho}} \label{eq:gen_i_def_2}
\end{multline}
where we use ``$\stackrel{\mathscr{L}}{=}$'' to denote equality in distribution, and where $\Gamma(\cdot)$ and $\tilde{\gamma}(\cdot)$ denote the gamma function and the regularized lower incomplete gamma function, respectively. In \eqref{eq:gen_i_def_2}, $\{\gamoneL\}_{\ell=1}^L$ are i.i.d. $\text{Gamma}(1,1)$-distributed random variables, and $\{\gamtwoL\}_{\ell=1}^L$ are i.i.d. $\text{Gamma}(\T-1,1)$-distributed random variables. For brevity, we let $i_{\ell,s}(\rho)\triangleq i_s(\Xk;\Yk)$ and denote the expectation and the variance of $i_{\ell,s}(\rho)$ by $I_s\lro{\rho} \triangleq \mathsf{E}\bigl[i_{\ell,s}(\rho)\bigr]$ and $V_s\lro{\rho}  \triangleq \mathsf{Var}\bigl[i_{\ell,s}(\rho)\bigr]$, respectively.
\subsection{Nonasymptotic Bounds}\label{sec:nonasym_spec}
We next present the nonasymptotic bounds that we shall use in this paper. As upper bound on $\epsilon^{*}(L,T,R,\rho)$, we use the \rcus{} bound~\cite[Th.~1]{Martinez_Guillen_i_Fabregas_2011}, which states that, for every $s> 0$, there exists a channel code of blocklength $LT$ and rate $R$ satisfying
\begin{IEEEeqnarray}{rCl} \label{eq:rcus_eps_2}
	\epsilon^*\lro{L,T,R,\rho} & \leq & \mathsf{P}\lrho{ \sum_{\ell=1}^L\lro{ I_s(\rho) -i_{\ell,s}(\rho)}\geq LI_s(\rho)+ \log U-L\T R}\IEEEeqnarraynumspace\label{eq:rcus_eps}
\end{IEEEeqnarray}
where $U$ is uniformly distributed on the interval $[0,1]$.

To present a lower bound on $\epsilon^*\lro{L,T,R,\rho}$, consider the auxiliary output pdf
\begin{equation}\label{eq:output_exp_ach}
\text{q}_{\mathbf{Y}_\ell}^s(\yk) \triangleq \frac{1}{\mu(s)} \left(\int \text{p}_{\Y_l|\X_l}(\yk|\xk)^s \mathrm{d} \bar{\text{P}}_{\mathbf{X}}(\xk) \right)^{1/s}, \quad s>0
\end{equation}
where $\mu(s)$ is a normalizing factor. We define the \mbox{\emph{generalized mismatched information density}}\footnote{We use the word ``mismatched'' to indicate that the auxiliary output pdf $\text{q}^s_{\mathbf{Y}_\ell}$ is not necessarily the one induced by the input distribution and the channel.}~as
\begin{IEEEeqnarray}{lCl}
	j_s(\Xk;\Yk)  &\triangleq&  \log \frac{\text{p}_{\Y_l|\X_l}(\Y_l|\X_l)}{\text{q}_{\mathbf{Y}_\ell}^s(\yk)} 	= \log \mu(s) + \frac{1}{s} i_s(\Xk;\Yk), \quad s>0. \label{eq:mis_inf_dens_summand}
\end{IEEEeqnarray}
For brevity, we define $j_{\ell,s}(\rho)\triangleq j_s(\Xk;\Yk)$ and $J_s(\rho)\triangleq \mathsf{E}\lrho{j_{\ell,s}(\rho)}.$
When $s=1$, we have $j_{\ell,1}(\rho) = i_{\ell,1}(\rho)$ and $ J_1(\rho) = I_1(\rho)$, in which case we omit the subscript and simply write $i_{\ell}(\rho)\triangleq i_{\ell,1}(\rho)$, $V(\rho)=V_1(\rho)$, and~$C(\rho) \triangleq I_1(\rho)$.\footnote{Recall that we chose the input distribution to be USTM, which is capacity achieving for the power constraint \eqref{power_const}.}

A lower bound on $\epsilon^{*}(L,T,R,\rho)$ follows by evaluating the MC bound \cite[Th. 31]{Polyanskiy_Poor_Verdu} for the auxiliary pdf $\text{q}_{\mathbf{Y}_\ell}^s$ and using~\cite[Eq.~(102)]{Polyanskiy_Poor_Verdu}. This yields that, for every channel code of blocklength $LT$ and rate $R$, we have
\begin{IEEEeqnarray}{lCl} 
	\epsilon^{*}(L,T,R,\rho) &\geq& \mathsf{P}\lrho{\sum_{\ell=1}^L\lro{I_s(\rho)-{i_{\ell,s}(\rho)}}\geq sLJ_s(\rho) - s\log\xi} - e^{\lro{\log\xi - L\T R}}\label{eq:MC_numerical}
\end{IEEEeqnarray}
for every $\xi>0$ and $s>0$.
\section{Saddlepoint Expansion}\label{sec:saddle_exp}
Let $Z_{1,\theta},\ldots,Z_{n,\theta}$ be a sequence of i.i.d., real-valued, zero-mean, random variables. The distribution of $Z_{\ell,\theta}$ depends on $\theta\in\Theta$, where $\Theta$ denotes the set of possible values of $\theta$.

For future reference, the moment generating function (MGF) of $Z_{\ell,\theta}$ is defined as
\begin{equation}\label{eq:MGF_x_def}
m_\theta(\zeta) \triangleq  \mathsf{E}\lrho{e^{\zeta Z_{\ell,\theta}}}
\end{equation}
and the CGF is defined as
\begin{equation}\label{eq:CGF_x_def}
\psi_\theta(\zeta) \triangleq \log m_\theta(\zeta).
\end{equation}
Furthermore, the characteristic function is defined as
\begin{equation}\label{eq:CH_x_def}
\varphi_\theta(\zeta) \triangleq  \mathsf{E}\lrho{e^{\imath\zeta Z_{\ell,\theta}}}
\end{equation}
where $\imath\triangleq\sqrt{-1}$. We denote by $m^{(k)}_\theta$ and $\psi^{(k)}_\theta$ the $k$-th derivative of $\zeta\mapsto m_\theta(\zeta)$ and $\zeta\mapsto \psi_\theta(\zeta)$, respectively.
For the first three derivatives we also use the notation $m_\theta'$, $m_\theta''$, $m_\theta'''$, $\psi_\theta'$, $\psi_\theta''$, and~$\psi_\theta'''$.

A random variable $Z_{\ell,\theta}$ is said to be \emph{lattice} if it is supported on the points $b$, $b\pm h$, $b \pm 2h$\dots for some $b$ and $h$. A random variable that is not lattice is said to be \emph{nonlattice}. It can be shown that a random variable is nonlattice if, and only if, \cite[Ch.~XV.1, Lemma~4]{Feller}
\begin{equation}
|\varphi_{\theta}(\zeta)| < 1, \quad \textnormal{for every $\zeta\neq 0$.}
\end{equation}
We shall say that a family of random variables $Z_{\ell,\theta}$ (parametrized by $\theta$) is nonlattice if
\begin{equation}
\sup_{\theta\in\Theta}\lrvo{\varphi_\theta(\zeta)}<1, \quad \textnormal{for every $\zeta\neq 0$.}
\end{equation}
Similarly, we shall say that a family of distributions (parametrized by $\theta$) is nonlattice if the corresponding family of random variables is nonlattice.

The following proposition presents saddlepoint expansions for families of random variables that are nonlattice. In Section \ref{sec:saddle_RBF}, these saddlepoint expansions will then be applied to the nonasymptotic bounds \eqref{eq:rcus_eps_2} and \eqref{eq:MC_numerical}. While similar expansions could also be derived for families of lattice random variables, this would require a separate proof. Thus, for the sake of compactness, and because for many channel models of interest---including the Rayleigh block-fading channel \eqref{ch_model}---the generalized information densities appearing in \eqref{eq:rcus_eps_2} and \eqref{eq:MC_numerical} are nonlattice, we do not consider lattice families of random variables in this paper.

\begin{proposition}\label{pr:main_saddle}
	Let the family of i.i.d.\ random variables $\lrbo{Z_{\ell,\theta}}_{\ell=1}^n$ (parametrized by $\theta$) be nonlattice. Suppose that there exists a $\zeta_0 > 0$ such that
	\begin{equation}\label{eq:MGF_unif_bounded}
		\sup_{\theta\in\Theta, |\zeta|<\zeta_0}\lrvo{m^{(4)}_\theta(\zeta)}<\infty
	\end{equation}
	and
	\begin{equation}\label{eq:inf_second_positive}
		\inf_{\theta\in\Theta,|\zeta|<\zeta_0}\psi_\theta''(\zeta)>0.
	\end{equation}
	Then, we have the following results:

	\emph{Part 1}:
	If for a given $\gamma\geq 0$ there exists a $\tau\in[0,\zeta_0)$ such that $n\psi_\theta'(\tau) = \gamma$,\footnote{In general, $\tau$ depends on $n$, $\theta$, and $\gamma$. For the sake of compactness, we do not make this dependence explicit in the notation.} then
	\begin{IEEEeqnarray}{lCl}\label{eq:saddlepoint}
		\mathsf{P}\lrho{\sum\limits_{\ell=1}^n Z_{\ell,\theta} \geq \gamma} &=&  e^{n[\psi_\theta(\tau)-\tau\psi_\theta'(\tau)]}\lrho{\Qexp_\theta\lro{\tau,\tau}+\frac{K_\theta(\tau,\tau,n)}{\sqrt{n}}+o\lro{\frac{1}{\sqrt{n}}}}
	\end{IEEEeqnarray}
	where
	\begin{subequations}
		\begin{IEEEeqnarray}{rCl}
			\Qexp_\theta\lro{u,\tau} & \triangleq & e^{n\frac{u^2}{2}\psi_{\theta}''(\tau)}Q\lro{u\sqrt{n\psi_{\theta}''(\tau)}}\label{eq:help_fcn_theta}\\
			K_\theta(u,\tau,n) & \triangleq & \frac{\psi_\theta'''(\tau)}{6\psi_\theta''(\tau)^{3/2}}\biggl(-\frac{1}{\sqrt{2\pi}}+\frac{u^2n\psi_\theta''(\tau)}{\sqrt{2\pi}} - u^3\psi_\theta''(\tau)^{3/2}n^{3/2}\Qexp_\theta(u,\tau)\biggr)\IEEEeqnarraynumspace\label{eq:K_def}
		\end{IEEEeqnarray}
	\end{subequations}
	and $o(1/\sqrt{n})$ comprises terms that vanish faster than $1/\sqrt{n}$ and are uniform in $\tau$ and $\theta$, i.e.,
  	\begin{equation}
		\lim_{n\to\infty} \sup_{\tau\in[0,\zeta_0), \theta\in\Theta} \frac{o(1/\sqrt{n})}{1/\sqrt{n}} = 0.
  	\end{equation}

	\emph{Part 2}:
	Let $U$ be uniformly distributed on $[0,1]$. If for a given $\gamma\geq0$ there exists a $\tau\in[0,\zeta_0)$ such that $n\psi_\theta'(\tau) = \gamma$, then
	\begin{multline}
		\mathsf{P}\lrho{\sum\limits_{\ell=1}^n Z_{\ell,\theta} \geq \gamma+\log U}
		\\= e^{n[\psi_\theta(\tau)-\tau\psi_\theta'(\tau)]}\lrho{ \Qexp_\theta(\tau,\tau) + \Qexp_\theta(1-\tau,\tau) +\frac{K_\theta(\tau,\tau,n)-K_\theta(1-\tau,\tau,n)}{\sqrt{n}}+o\lro{\frac{1}{\sqrt{n}}}}\label{eq:saddlepoint_U}
	\end{multline}
where $o(1/\sqrt{n})$ is uniform in $\tau$ and $\theta$.
\end{proposition}
\begin{IEEEproof}
	Part~1 is proved in Appendix~\ref{app:proof_propMC} and Part~2 is proved in Appendix~\ref{app:proof_propRCUs}.
\end{IEEEproof}
\begin{remark}\label{rm:CGF_finite}
A Taylor series expansion of $\zeta\mapsto m^{(k)}_\theta(\zeta)$ around zero demonstrates that, if $m^{(k)}_\theta(\zeta)$ is bounded in $\theta\in\Theta$ and $|\zeta|<\zeta_0$, then the same is also true for $m^{(k-1)}_\theta(\zeta)$. Consequently, \eqref{eq:MGF_unif_bounded} implies that $\zeta\mapsto m_\theta(\zeta)$ and its first four derivatives are bounded in $\theta\in\Theta$ and $|\zeta|<\zeta_0$. This in turn implies that $\zeta\mapsto m_\theta(\zeta)$ is analytic on $\tau\in(-\zeta_0,\zeta_0)$ for every $\theta\in\Theta$. Since $Z_\ell$ has zero mean by assumption, we further have that $m_\theta(\zeta)\geq 1$ by Jensen's inequality. We conclude that all derivatives of $\zeta\mapsto\psi_\theta(\zeta)$ exist on $\tau\in(-\zeta_0,\zeta_0)$ for every $\theta\in\Theta$, and
	\begin{equation}\label{eq:CGF_unif_bounded}
	\sup_{\theta\in\Theta, |\zeta|<\zeta_0}\bigl|\psi^{(k)}_\theta(\zeta)\bigr|<\infty, \quad k=0,\ldots,4.
	\end{equation}
\end{remark} 
\begin{remark} The asymptotic behaviors of $\Qexp_{\theta}(u,\tau)$ and $K_\theta(u,\tau,n)$ depend critically on the asymptotic behavior of $u$. Indeed, the $\tau$ satisfying $n\psi_{\theta}'(\tau)=\gamma$, and hence also $u\in\{\tau,1-\tau\}$, depends in general on $n$. The term $\Qexp_{\theta}(u,\tau)$ converges to a constant when $u$ decays at least like $1/\sqrt{n}$, and it is of order $n^{a-1/2}$ when $u$ is of order $n^{-a}$, $0\leq a <1/2$. Similarly, $K_{\theta}(u,\tau,n)$ converges to a constant when $u$ decays at least like $1/\sqrt{n}$, and it vanishes as $n\to\infty$ when $u$ is of order $n^{-a}$, $0\leq a < 1/2$.
\end{remark}

Since the difference $K_\theta(\tau,\tau,n)-K_\theta(1-\tau,\tau,n)$ is difficult to evaluate, in the following corollary we present an upper bound on the saddlepoint expansion \eqref{eq:saddlepoint_U} that is easier to evaluate.

\begin{corollary}\label{co:K_RCUs}
	Assume that there exists a $\zeta_0>0$ satisfying \eqref{eq:MGF_unif_bounded} and \eqref{eq:inf_second_positive}. If for a given $\gamma\geq 0$ there exists a $\tau\in[0,\min\{\zeta_0,1-\delta\})$ (for some arbitrary $\delta>0$ independent of $n$ and $\theta$) such that $n\psi_\theta'(\tau) = \gamma$, then the saddlepoint expansion \eqref{eq:saddlepoint_U} can be upper-bounded as
	\begin{IEEEeqnarray}{lCl}
		\mathsf{P}\lrho{\sum\limits_{\ell=1}^n Z_{\ell,\theta} \geq \gamma+\log U}
		&&\leq e^{n[\psi_\theta(\tau)-\tau\psi_\theta'(\tau)]}\lrho{\Qexp_\theta(\tau,\tau)+\Qexp_\theta(1-\tau,\tau)+\frac{\hat{K}_\theta(\tau)}{\sqrt{n}}+o\lro{\frac{1}{\sqrt{n}}}}\IEEEeqnarraynumspace\label{eq:saddlepoint_U_upper}
	\end{IEEEeqnarray}
	where
	\begin{equation}\label{eq:K_up_rcus}
	\hat{K}_\theta(\tau) \triangleq \frac{1}{\sqrt{2\pi}}\frac{\psi_\theta'''(\tau)}{6\psi_\theta''(\tau)^{3/2}}
	\end{equation}
	and $o(1/\sqrt{n})$ is uniform in $\tau$ and $\theta$.
\end{corollary}
\begin{IEEEproof}
	See Appendix~\ref{app:proof_coRCUs}.
\end{IEEEproof}
\section{Saddlepoint Expansions of \rcus{} and MC Bounds}\label{sec:saddle_RBF}
We next apply Proposition~\ref{pr:main_saddle} and Corollary~\ref{co:K_RCUs} to the \rcus{} bound \eqref{eq:rcus_eps} and MC bound \eqref{eq:MC_numerical} to obtain their saddlepoint expansions. To this end, we first express the MGF and the CGF in terms of the generalized information density \eqref{eq:inf_dens_summand_def} and discuss their regions of convergence.

The MGF and the CGF of $I_s(\rho)-i_{\ell,s}(\rho)$ are defined as
\begin{IEEEeqnarray}{lCl}
m_{\rho,s}(\tau) & \triangleq & \mathsf{E}\lrho{e^{\tau \lro{I_s(\rho)-i_{\ell,s}(\rho)}}} \label{eq:MGF_I_i}\\
\psi_{\rho,s}(\tau) & \triangleq & \log m_{\rho,s}(\tau) \label{eq:CGF_I_i}
\end{IEEEeqnarray}
and depend on the parameters $\theta = \{\rho,s\}$. For some arbitrary $0<\underline{s}<\overline{s}<\infty$, $0<\underline{\rho}<\overline{\rho}<\infty$, $0<a<1$, and $0<b<\min\lrbo{\frac{T}{T-1},\frac{1+T\overline{\rho}}{\overline{s}T\overline{\rho} }}$, let
\begin{equation}
\mathcal{S} \triangleq \{(\tau,\rho,s)\in\mathbb{R}^3\colon \tau\in[a,b], \rho\in[\underline{\rho},\overline{\rho}], s\in[\underline{s},\overline{s}]\}.
\end{equation}
It can be shown that \cite[Lemma~4.2]{LanchoPhd}
\begin{equation}
\sup\limits_{(\tau,\rho,s)\in\mathcal{S}}m_{\rho,s}^{(k)}(\tau) < \infty\label{eq:MGF_original_lm_p2}
\end{equation}
for every nonnegative integer $k$. So $\mathcal{S}$ is in the region of convergence of $m_{\rho,s}$. We are now ready to present the following saddlepoint expansions.

\begin{theorem}[Saddlepoint Expansion \rcus{}]\label{thm:RCUs_SP}
The coding rate $R$ and minimum error probability~$\epsilon^*$ can be parametrized by $(\tau,\rho,s)\in\mathcal{S}$ as
\begin{subequations}
\begin{IEEEeqnarray}{rCl}
R(\tau,s) &=&\frac{1}{\T}(I_s(\rho)-\psi_{\rho,s}'(\tau))\label{eq:rate_SP_rcus}   \IEEEeqnarraynumspace\\
	\epsilon^{*}(\tau,s)  &\leq &   e^{L\lrho{\psi_{\rho,s}(\tau)- \tau \psi_{\rho,s}'(\tau)}} \biggl[\Qexp_{\rho,s}\lro{\tau,\tau} + \Qexp_{\rho,s}\lro{1-\tau,\tau} + \frac{\hat{K}_{\rho,s}(\tau)}{\sqrt{L}} + o\lro{\frac{1}{\sqrt{L}}}\biggr]\IEEEeqnarraynumspace\label{eq:saddle_RCUs_numeric_gen}
\end{IEEEeqnarray}
\end{subequations}
where  
$o(1/\sqrt{L})$ is uniform in $\tau$, $s$, and~$\rho$.\footnote{The error terms appearing in the asymptotic expansions depend, in general, also on $T$. However, we do not make this dependence explicit in the notation, since we view $T$ as a fixed parameter.}
\end{theorem}
\begin{IEEEproof}
The desired result follows by applying Corollary~\ref{co:K_RCUs} to \eqref{eq:rcus_eps}.  Indeed, it can be shown that the family of random variables $I_s(\rho)-i_{s,\ell}(\rho)$ (parametrized by $(\rho,s)$) is nonlattice \cite[Lemma~B.2]{LanchoPhd}. Furthermore, \eqref{eq:MGF_original_lm_p2} implies that the first condition~\eqref{eq:MGF_unif_bounded} required for Proposition~\ref{pr:main_saddle} and Corollary~\ref{co:K_RCUs} is satisfied. Regarding the second condition \eqref{eq:inf_second_positive}, it can be observed that $V_s(\rho)$ is strictly increasing in $\rho$ (for a fixed $s$) and strictly increasing in $s$ (for a fixed $\rho$). Consequently, it is bounded away from zero for every $\rho\geq\underline{\rho}$ and $s\geq \underline{s}$ (for arbitrary $\underline{\rho}>0$ and $\underline{s}>0$). Since $\psi_{\rho,s}''(0) = V_s\lro{\rho}$, it follows from Lemma~\ref{lm:CGF2_bounded_away} in Appendix~\ref{app:CGF2_boundedaway} that the second condition \eqref{eq:inf_second_positive} required in Proposition~\ref{pr:main_saddle} and Corollary~\ref{co:K_RCUs} is satisfied, too.
		\end{IEEEproof}
\begin{remark}
The set $\mathcal{S}$ with $\overline{s}=1$ includes $0 \leq \tau < 1$. In this case, the identity \eqref{eq:rate_SP_rcus} characterizes all rates $R$ between the critical rate, defined as \cite[Eq. (5.6.30)]{Gallager}
\begin{IEEEeqnarray}{rCl}
\Rcr{\rho} \triangleq \frac{1}{\T}\lro{I_s(\rho) - \psi_{\rho,s}'(1)} \label{eq:Rcr_E0}
\end{IEEEeqnarray}
and $I_s(\rho)$. Solving \eqref{eq:rate_SP_rcus} for $\tau$, we obtain from Theorem~\ref{thm:RCUs_SP} an upper bound on the minimum error probability $\epsilon^{*}(L,T,R,\rho)$ as a function of the rate $R\in(\Rcr{\rho},I_s(\rho)]$, $s\in(0,1]$.
\end{remark}

\begin{theorem}[Saddlepoint Expansion MC]\label{thm:MC_Saddle}
For every rate $R$ and $(\tau,\rho,s)\in\mathcal{S}$
	\begin{equation}
		\epsilon^{*}(L,T,R,\rho) \geq - e^{L\lrho{J_s(\rho)-\frac{\psi_{\rho,s}'(\tau)}{s} - \T R}} + e^{L\lrho{\psi_{\rho,s}(\tau)- \tau \psi_{\rho,s}'(\tau)}} \lrho{ \Qexp_{\rho,s}\lro{\tau,\tau}+ \frac{K_{\rho,s}(\tau,\tau,L)}{\sqrt{L}} + o\lro{\frac{1}{\sqrt{L}}}} \label{eq:MC_Saddle}
	\end{equation}
	where 
	the $o(1/\sqrt{L})$ term is uniform in $\tau$, $s$, and $\rho$.
\end{theorem}
\begin{IEEEproof}
The inequality \eqref{eq:MC_Saddle} follows by applying Proposition~\ref{pr:main_saddle} to \eqref{eq:MC_numerical} with $\log\xi=LJ_s(\rho)-L\psi_{\rho,s}'(\tau)/s$. The rest of the proof is similar to the proof of Theorem~\ref{thm:RCUs_SP}. 
\end{IEEEproof}
\section{Normal Approximations and Error Exponents}\label{sec:NA_err}
It is possible to recover the normal approximation and the error exponent of the channel from the saddlepoint expansions. Detailed derivations can be found in \cite[Chs.~6.3--6.4]{LanchoPhd}.

Specifically, an asymptotic analysis of \eqref{eq:rate_SP_rcus} and \eqref{eq:MC_Saddle} yields the normal approximation
\begin{equation}\label{eq:NA}
R^*(L,T,\epsilon,\rho) = \frac{C(\rho)}{T} - \sqrt{\frac{V(\rho)}{LT^2}}Q^{-1}\lro{\epsilon} + \mathcal{O}\lro{\frac{\log L}{L}}
\end{equation}
where $\mathcal{O}\bigl((\log L)/L\bigr)$ comprises terms that are of order of $(\log L) / L$ and are uniform in $\rho$.

Note that there are no closed-form expressions for $C(\rho)$ and $V(\rho)$, so these quantities must be evaluated numerically by computing the mean and variance of $i_{\ell}(\rho)$. A closed-form high-SNR normal approximation was presented in \cite[Th. 2]{LKD17}. This expansion can be recovered from \eqref{eq:NA} by using that \cite[Eqs.~(38) and (39)]{LKD17}
\begin{IEEEeqnarray}{rCl}
	C(\rho) & = & (T-1)\log(T\rho)-\log\Gamma(T) -(T-1)\lrho{\log(1+T\rho) + \> \frac{T\rho}{1+T\rho} - \psi(T-1) }\nonumber\\
	&& {} +{_2F_1\lro{1,T-1;T;\frac{T\rho}{1+T\rho}}} + o_{\rho}(1)\IEEEyesnumber\IEEEyessubnumber*\label{eq:sec:first_first_moment_closed}\\
V(\rho) & = & (T-1)^2\frac{\pi^2}{6}+(T-1) + o_{\rho}(1) \label{eq:second_second_moment_high_SNR}
\end{IEEEeqnarray}
where $o_{\rho}(1)$ comprises terms that are uniform in $L$ and vanish as $\rho\to\infty$. In \eqref{eq:sec:first_first_moment_closed}, $\psi(\cdot)$ denotes the digamma function and $_2F_1(\cdot,\cdot;\cdot;\cdot)$ denotes the Gaussian hypergeometric function.

The saddlepoint expansions \eqref{eq:saddle_RCUs_numeric_gen} and \eqref{eq:MC_Saddle} can be also written as an exponential term times a subexponential factor. The exponential terms of both expansions coincide for rates $\Rcrb{\rho}<R<C(\rho)$, where $\Rcrb{\rho}$ is the critical rate \eqref{eq:Rcr_E0} evaluated at $s=1/2$. So they characterize the \textit{reliability function}, defined as
\begin{equation}
	E_r(\T,R,\rho) \triangleq \lim\limits_{L\to\infty} -\frac{1}{L}\log \epsilon^*(L,T,R,\rho).
\end{equation}
Let now $\underline{\rho}\leq\rho\leq\overline{\rho}$ and $\underline{\tau}<\tau<\overline{\tau}$ for some arbitrary $0<\underline{\rho}<\overline{\rho}<\infty$ and $0<\underline{\tau}<\overline{\tau}<1$, and set $s_{\tau}\triangleq 1/(1+\tau)$. The coding rate $R$ and the minimum error probability $\epsilon^*$ can be parametrized by $\tau\in(\underline{\tau},\overline{\tau})$ as 
\begin{IEEEeqnarray}{rCl}\label{eq:err_exp_approx_def}
	R(\tau) &=& \frac{1}{T}\left(I_{s_{\tau}}(\rho) - \psi_{\rho,s_{\tau}}'(\tau)\right)\IEEEyesnumber\IEEEyessubnumber*\label{eq:tau_star_err_exp_def}\\
	\underline{\mathsf{A}}_{\rho}(\tau) &\leq& \epsilon^*(L,T,R,\rho)e^{-L\lrho{\psi_{\rho,s_{\tau}}(\tau)-\tau \psi_{\rho,s_{\tau}}'(\tau)}} \leq \overline{\mathsf{A}}_{\rho}(\tau) \IEEEeqnarraynumspace \label{eq:rcus_SP_cor_2_th}
\end{IEEEeqnarray}
where
\begin{IEEEeqnarray}{lCl}\label{eq:preexp}
	\overline{\mathsf{A}}_{\rho}(\tau) & \triangleq & \frac{1}{\sqrt{2\pi L \tau^2 \psi_{\rho,s_\tau}''(\tau)}} + \frac{|\hat{K}_{\rho,s_\tau}(\tau)|}{\sqrt{L}} + \frac{1}{\sqrt{2\pi L(1-\tau)^2  \psi_{\rho,s_\tau}''(\tau)}} +o\lro{\frac{1}{\sqrt{L}}} \IEEEeqnarraynumspace\IEEEyesnumber\IEEEyessubnumber* \label{eq:preexp1}\\
	 \underline{\mathsf{A}}_{\rho}(\tau) & \triangleq & \frac{s_\tau^{\frac{1}{s_\tau}}}{\tau\lro{2\pi L \psi_{\rho,s_\tau}''(\tau)}^{\frac{1}{2s_\tau}}} + o\lro{\frac{1}{L^{\frac{1}{2s_\tau}}}}. \label{eq:preexp2}
\end{IEEEeqnarray}
The little-$o$ term in \eqref{eq:preexp1} is uniform in $\rho$ and $\tau$. The little-$o$ term in \eqref{eq:preexp2} is uniform in $\rho$ (for every given $\tau$).

The products of $\overline{\mathsf{A}}_{\rho}(\tau)$ and $\underline{\mathsf{A}}_{\rho}(\tau)$ with $e^{L\lrho{\psi_{\rho,s_{\tau}}(\tau)-\tau \psi_{\rho,s_{\tau}}'(\tau)}}$ yield approximations of the \rcus{} and MC saddlepoint bounds, respectively. Note that the first term of $ \underline{\mathsf{A}}_{\rho}(\tau)$ is positive and of order $L^{-\frac{1+\tau}{2}}$. This order coincides with that of the subexponential factor of the random-coding upper bound on the error probability for some memoryless channels and rates above the critical rate \cite{Altug2014,Honda2018}.
 It follows from \eqref{eq:tau_star_err_exp_def}, \eqref{eq:rcus_SP_cor_2_th}, and the behaviors of $\overline{\mathsf{A}}_{\rho}(\tau)$ and $\underline{\mathsf{A}}_{\rho}(\tau)$ that the reliability function $E_r(\T,R,\rho)$ can be parametrized by $\tau\in(0,1)$~as
\begin{IEEEeqnarray}{rCl}
	E_r(\tau) & = & \tau \psi_{\rho,\frac{1}{1+\tau}}'(\tau) - \psi_{\rho,\frac{1}{1+\tau}}(\tau)\IEEEyesnumber\IEEEyessubnumber* \label{eq:rel_func_sol_th} \\
	R(\tau) & = & \frac{1}{T}\lro{I_{\frac{1}{1+\tau}}(\rho) - \psi_{\rho,\frac{1}{1+\tau}}'(\tau)}. \label{eq:rel_func_Rtau}
\end{IEEEeqnarray}		
\section{Numerical Examples}\label{sec:num_results}
To show the accuracy of the saddlepoint approximations, we evaluate them for different channel parameters and compare them with the nonasymptotic bounds as well as the refined asymptotic expansions available in the literature.

In all figures, we plot approximations of the \rcus{} bound in red and approximations of the MC bound in blue, which are obtained by disregarding the $o(1/\sqrt{L})$ terms in the saddlepoint expansions and by optimizing numerically over the parameters $s$ and $\tau$.  Solid lines (``saddlepoint") depict the saddlepoint approximations \eqref{eq:saddle_RCUs_numeric_gen} and~\eqref{eq:MC_Saddle}, and dashed lines (``pref+EE") depict \eqref{eq:rcus_SP_cor_2_th}. We further plot the nonasymptotic bounds \eqref{eq:rcus_eps} and~\eqref{eq:MC_numerical}, evaluated using Monte-Carlo simulations, with dots and highlight in grey the area where the true maximum coding rate lies. Finally, we plot the normal approximation~\eqref{eq:NA} (``NA") and the error-exponent approximations (``EEA" and ``$L^{-\frac{1+\tau}{2}}$+EEA'') that follow by solving
\begin{subequations}
\begin{IEEEeqnarray}{lCl}
\epsilon^*(L,T,R,\rho) & \approx & \exp\{-L E_r(\T,R,\rho)\}\label{eq:err_approx} \\
\epsilon^*(L,T,R,\rho) & \approx & \frac{1}{L^{\frac{1+\tau}{2}}}\exp\{-L E_r(\T,R,\rho)\} \label{eq:err_approx_2}
\end{IEEEeqnarray}
\end{subequations}
for $R$, where $\tau$ in \eqref{eq:err_approx_2} is the parameter for which \eqref{eq:rel_func_Rtau} is equal to $R$.

\begin{figure}[t!]
\centering
\begin{minipage}{0.48\textwidth}
	\centering
	\includegraphics[width=\columnwidth]{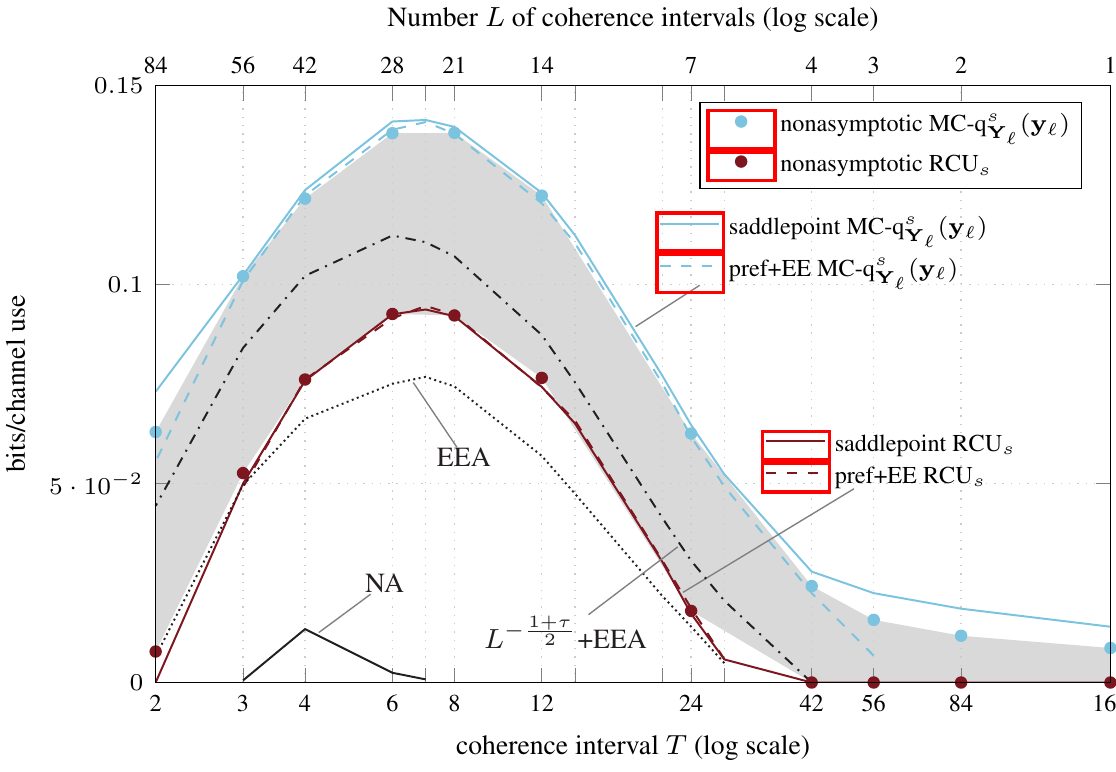}
	\vspace{-2.5em}
	\caption{$n=168$, $\epsilon=10^{-5}$, and $\rho=0$~dB.}
	\label{fig:trade_off_p0dB_n168_eps5}
\end{minipage}
\hfill
\begin{minipage}{0.48\textwidth}
	\centering
	\includegraphics[width=\columnwidth]{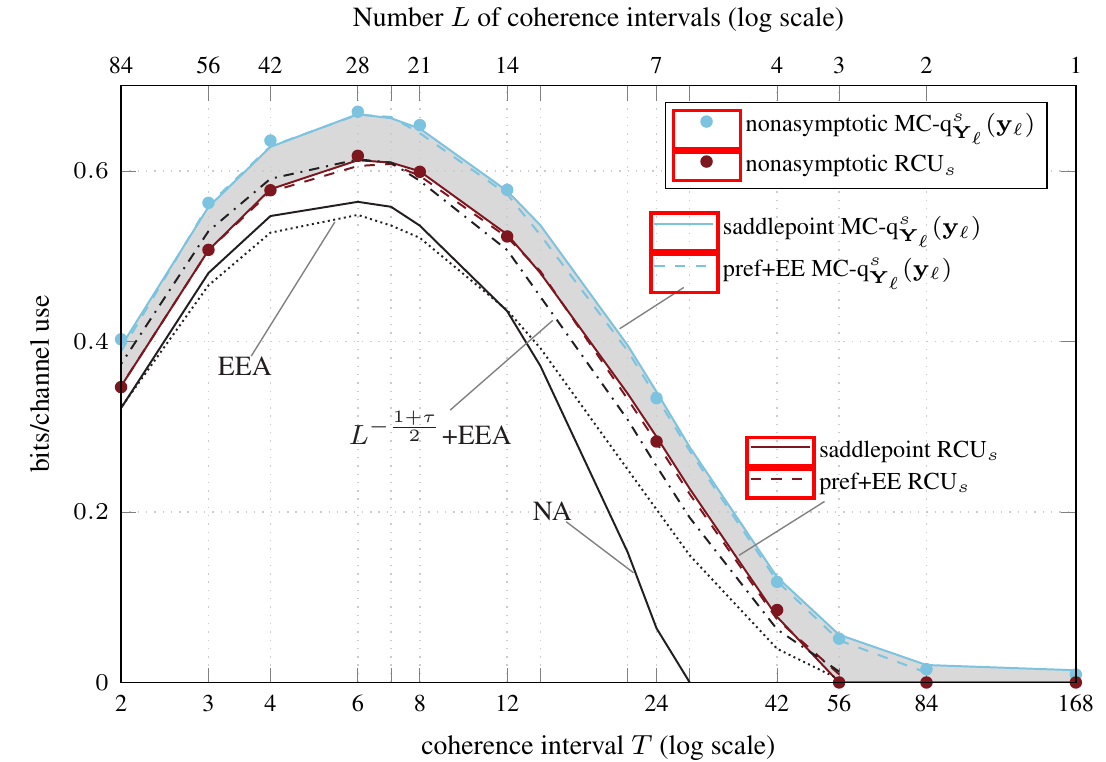}
	\vspace{-2.5em}
	\caption{$n=168$, $\epsilon=10^{-5}$, and $\rho=6$~dB.}
	\label{fig:trade_off_p6dB_n168_eps5}
\end{minipage}
\end{figure}

In Figs.~\ref{fig:trade_off_p0dB_n168_eps5} and~\ref{fig:trade_off_p6dB_n168_eps5}, we study $R^{*}(L,T,\epsilon,\rho)$ as a function of $L$ for $n=L\T=168$ (hence $\T$ is inversely proportional to $L$), $\epsilon=10^{-5}$, and SNR values $0$~dB and $6$~dB, respectively.  Observe that the approximations \eqref{eq:saddle_RCUs_numeric_gen}, \eqref{eq:MC_Saddle}, and \eqref{eq:rcus_SP_cor_2_th} are almost indistinguishable from the nonasymptotic bounds. Further observe that, compared to the saddlepoint expansions, the normal approximations and the error-exponent approximations are loose, although their accuracy increases for larger SNR values (and larger values of L, for the normal approximation). 
Finally, the error-exponent approximation ``$L^{-\frac{1+\tau}{2}}$+EEA" is accurate for the entire range of parameters.

\begin{figure}[t!]
\centering
\begin{minipage}{0.48\textwidth}
\centering
\includegraphics[width=\linewidth]{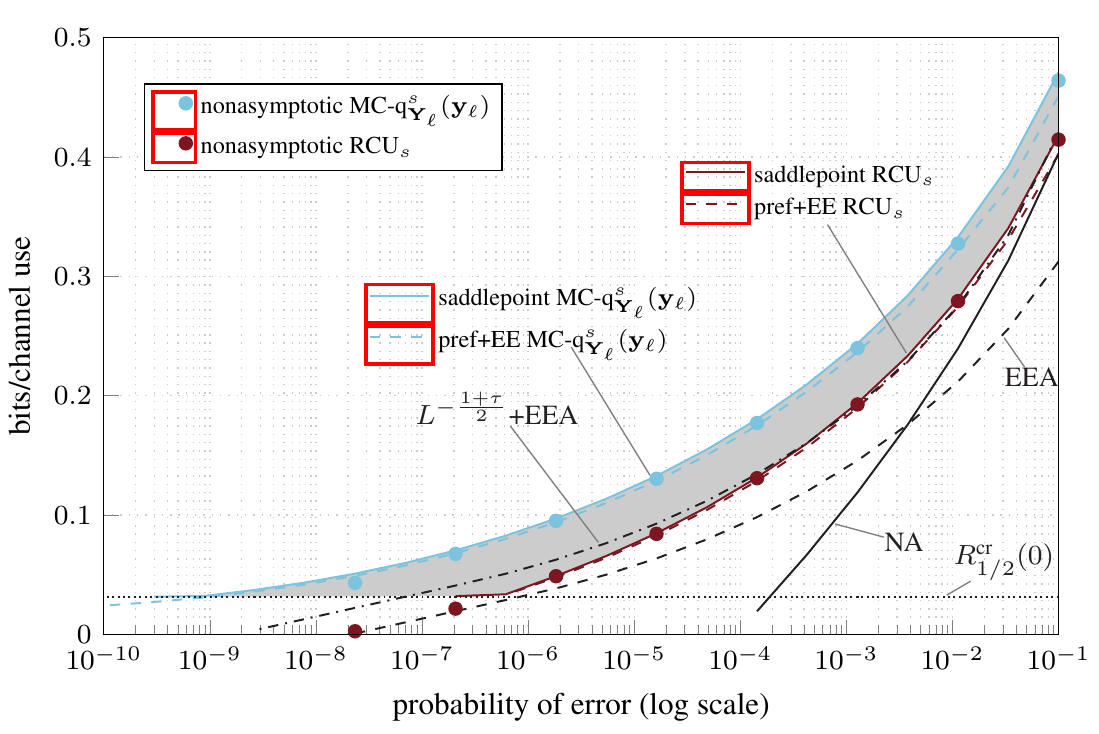}
\vspace{-2.5em}
\caption{$L=14$, $\T=12$, and $\rho=0$~dB.}
\label{fig:rate_vs_Pe_0db}
\end{minipage}%
\hfill
\begin{minipage}{0.48\textwidth}
\centering
\includegraphics[width=\linewidth]{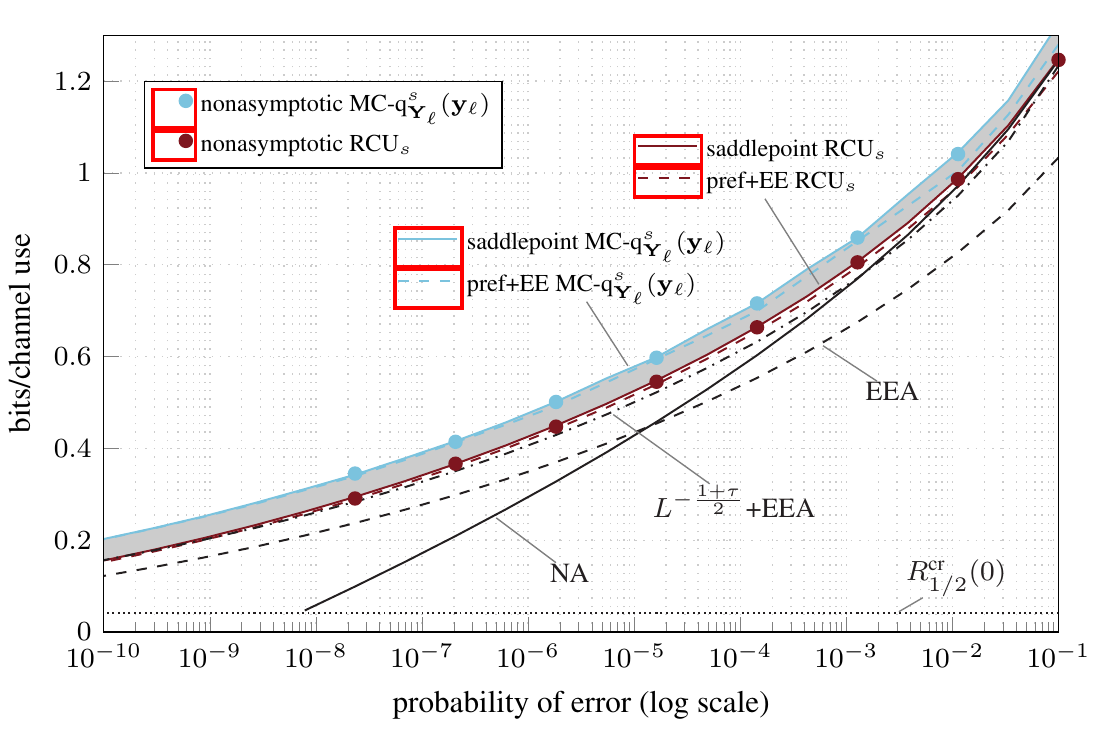}
\vspace{-2.5em}
\caption{$L=14$, $\T=12$, and $\rho=6$~dB.}
\label{fig:rate_vs_Pe_6db}
\end{minipage}
\end{figure}

In Figs.~\ref{fig:rate_vs_Pe_0db} and~\ref{fig:rate_vs_Pe_6db}, we study $R^{*}(L,T,\epsilon,\rho)$ as a function of $\epsilon$ for $n=168$ ($T=12$ and $L=14$) and SNR values $0$~dB and $6$~dB, respectively.
In addition to the aforementioned bounds and approximations, we plot $R_{1/2}^{\text{cr}}(0)$ to indicate the minimum rate that can be characterized by \eqref{eq:rate_SP_rcus}.
Observe that the approximations \eqref{eq:saddle_RCUs_numeric_gen}, \eqref{eq:MC_Saddle}, and \eqref{eq:rcus_SP_cor_2_th} are almost indistinguishable from the nonasymptotic bounds. Further observe how the normal approximation ``NA" becomes accurate for large error probabilities, whereas the error-exponent approximation ``EEA" becomes accurate for small error probabilities. In contrast, the error-exponent approximation ``$L^{-\frac{1+\tau}{2}}$+EEA" is accurate for the entire range of parameters.
\begin{figure}[t!]
\centering
\begin{minipage}{.48\textwidth}
\centering
\includegraphics[width=\linewidth]{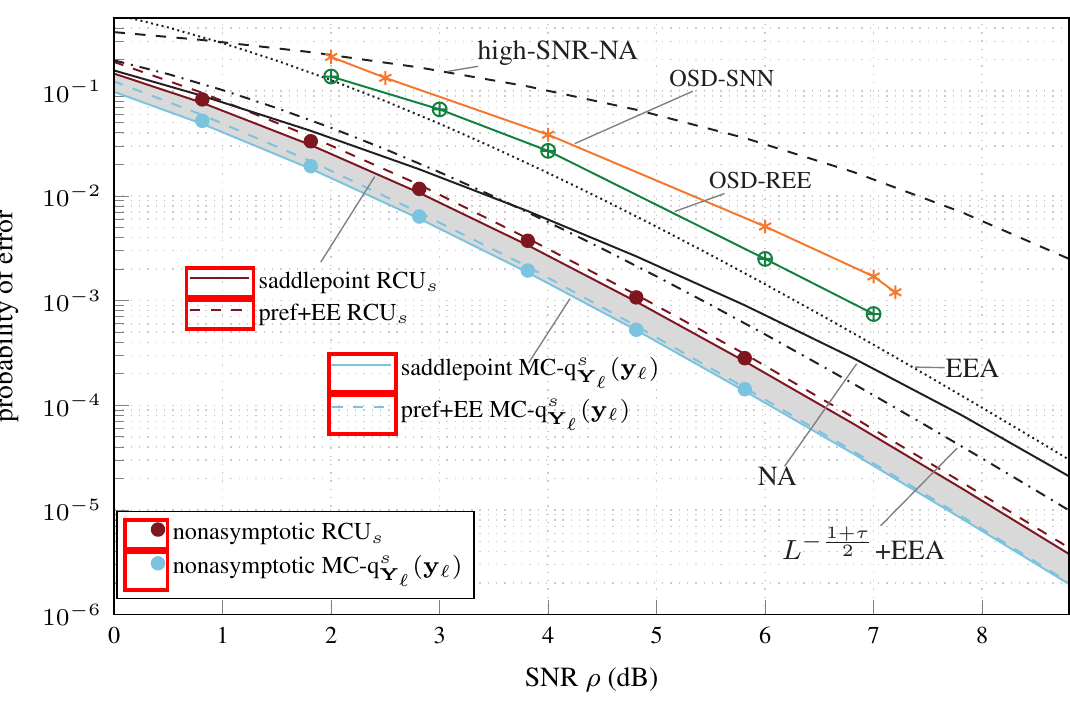}
\vspace{-2.5em}
\caption{$L=7$, $\T=24$, and $R=0.48$.}  
\label{fig:SNR_vs_Pe_1}
\end{minipage}%
\hfill
\begin{minipage}{.48\textwidth}
\centering
\includegraphics[width=\linewidth]{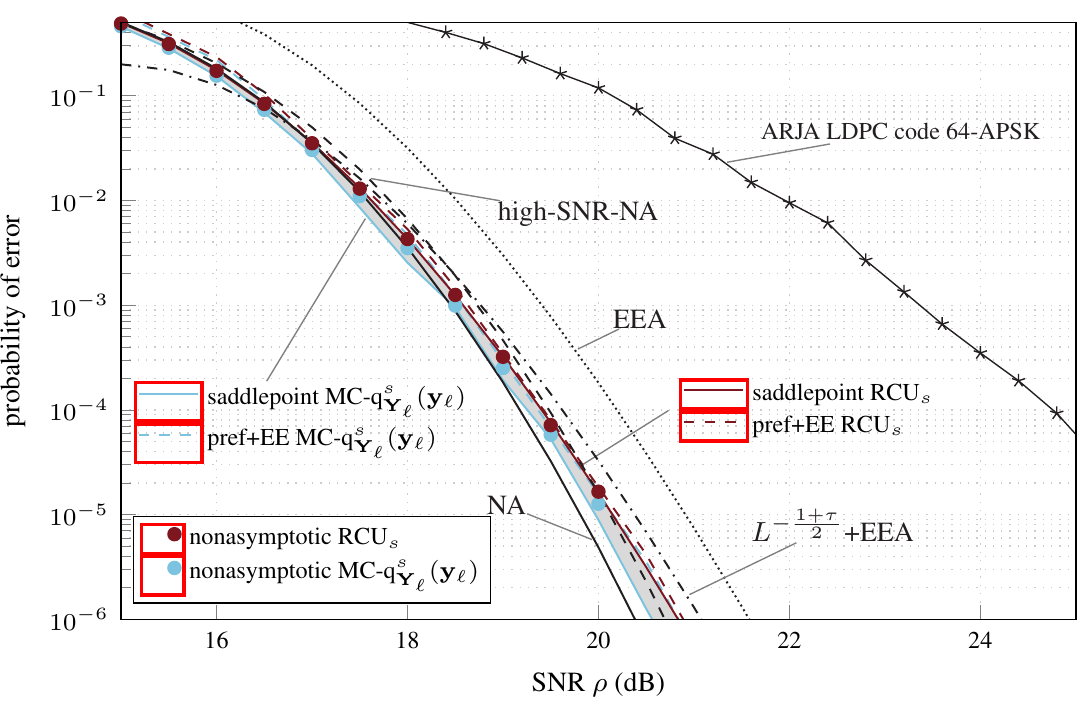}
\vspace{-2.5em}
\caption{$L=25$, $\T=20$, and $R=4$.}
\label{fig:SNR_vs_Pe_2}
\end{minipage}
\end{figure}

In Figs.~\ref{fig:SNR_vs_Pe_1} and \ref{fig:SNR_vs_Pe_2}, we study $\epsilon^{*}(L,T,R,\rho)$ for a fixed rate $R$ as a function of the SNR~$\rho$. Specifically, in Fig.~\ref{fig:SNR_vs_Pe_1} we show $\epsilon^{*}(L,T,R,\rho)$ for $n=168$ ($\T=24$ and $L=7$) and $R=0.48$, and in Fig.~\ref{fig:SNR_vs_Pe_2} we show $\epsilon^{*}(L,T,R,\rho)$ for $n=500$ ($\T=20$ and $L=25$) and $R=4$.
In both figures, we also show the high-SNR normal approximation (``high-SNR-NA'') derived in \cite[Th. 2]{LKD17}.
In addition to the aforementioned bounds and approximations, in Fig.~\ref{fig:SNR_vs_Pe_1} we show the simulated performance of a transmission scheme for which $4$ channel uses per coherence interval are allocated to coded pilot symbols belonging to a quaternary phase shift keying (QPSK) constellation, and the rest carry QPSK symbols encoded by a binary quasi-cyclic code. Decoding is performed via ordered statistics decoding (OSD), which uses scaled nearest-neighbor decoding (SNN) as the metric to solve the likelihood ratios. This coding scheme is denoted as ``OSD-SNN'' and is depicted in orange. We also show the simulated performance of a transmission scheme based on OSD-SNN that additionally performs a re-estimation of the fading coefficients based on the initial OSD decision. This coding scheme is denoted as ``OSD-REE'' and is depicted in green. The derivation and design of these codes can be found in~\cite{Ostman_2018}. Observe that, for this scenario, the error-exponent approximation ``$L^{-\frac{1+\tau}{2}}$+EEA" is not as accurate as the saddlepoint approximations. In Fig.~\ref{fig:SNR_vs_Pe_2} we also show the performance of an accumulate-repeat-jagged-accumulate (ARJA) low-density parity-check (LDPC) code combined with 64-APSK modulation, pilot-assisted channel estimation (2 pilot symbols per coherence block), and maximum likelihood channel estimation followed by mismatched nearest-neighbor decoding at the receiver (``ARJA LDPC code 64-APSK''); for details see \cite[Sec. 4]{Mustafa2019}. Observe that the gap between the performance of the presented transmission schemes and the rest of the curves is substantial in both figures. This suggests that more sophisticated (possibly joint) channel-estimation and decoding procedures together with shaping techniques need to be adopted to close the gap.
\section{Discussion}\label{sec:discussion}
In the following, we discuss the complexity of the numerical evaluation of the proposed saddlepoint approximations and compare it with that of the nonasymptotic bounds as well as of the refined asymptotic expansions available in the literature. We also provide additional remarks on when one should use which approximation.  
\subsection{Complexity Analysis}\label{sec:complex}
\begin{table*}
\caption{Complexity and computational time of the presented bounds and approximations for \newline $L=30$, $T=10$, $\epsilon=10^{-6}$, $\rho=6$~\textrm{dB}.}
\label{tb:complex_vs_time}
\centering
\begin{tabular}{|l || l | l |}
  \hline
\textbf{Bounds / Approximations} & \textbf{Comp.\ Complexity} & \textbf{Comp.\ Time}\\  \hline\hline
 & \rcus{}: $K N^{2L+1}$ (optimization over $s$) & \rcus{}: 3720~s\\ 
\raisebox{2ex}[-2ex]{nonasymptotic bounds} & MC: $K^2 N^{2L}$ (optimization over $s,\xi$) & MC: 3280~s\\\hline
 & & \rcus{}: 162~s \\
\raisebox{2ex}[-2ex]{saddlepoint approximations} &\raisebox{2ex}[-2ex]{$5 K^2 N^2$ (optimization over $\tau$, $s$)}  & MC: 194~s \\\hline
 &  & \rcus{}: 170~s \\
\raisebox{2ex}[-2ex]{prefactor-and-error-exponent approximations} & \raisebox{2ex}[-2ex]{$5 K N^2$ (optimization over $\tau$)} & MC: 207~s \\  \hline
error-exponent approximations & $3 K N^2$ (optimization over $\tau$) & 140~s \\ \hline
normal approximation & $2N^2$ (no optimization) & 0.7~s\\ \hline
high-SNR normal approximation & available in closed form & 5~s\\ \hline
\end{tabular}
\end{table*}

We denote by $N$ the cost of numerically evaluating a one-dimensional integral and by $K$ the cost of performing an optimization over an auxiliary parameter. In the following, we provide a coarse estimate of the complexity of the presented bounds and approximations in terms of $N$ and~$K$. The nonasymptotic bounds \eqref{eq:rcus_eps_2} and \eqref{eq:MC_numerical} require the evaluation of the distribution function of $\sum_{\ell=1}^L i_{\ell,s}(\rho)$. By \eqref{eq:gen_i_def_2}, the generalized information density $i_{\ell,s}(\rho)$ can be expressed in terms of the two random variables $\gamoneL$ and $\gamtwoL$. It follows that the evaluation of the \rcus{} bound requires the computation of a $(2L+1)$-dimensional integral (over $\{\gamoneL\}_{\ell=1}^L$, $\{\gamtwoL\}_{\ell=1}^L$, and $U$) and an optimization over $s$. Similarly,  the MC bound requires the computation of a $(2L)$-dimensional integral and an optimization over $s$ and $\xi$. The numerical evaluation of an $L$-dimensional integral has a complexity of roughly $N^L$. Hence, the overall complexity of the \rcus{} bound is $K N^{2L+1}$, and the complexity of the MC bound is $K^2 N^{2L}$. The saddlepoint approximations \eqref{eq:saddle_RCUs_numeric_gen} and \eqref{eq:MC_Saddle}, and the prefactor-and-error-exponent approximations \eqref{eq:rcus_SP_cor_2_th}, depend on $I_s(\rho)$, $\psi_{\rho,s}$, $\psi_{\rho,s}'$, $\psi_{\rho,s}''$, and $\psi_{\rho,s}'''$, so they can be obtained by solving 5 two-dimensional integrals and by optimizing over $(\tau,s)$ (saddlepoint approximations) or over $\tau$ (prefactor-and-error-exponent approximations). The error-exponent approximations \eqref{eq:err_approx}--\eqref{eq:err_approx_2} can be obtained by evaluating $I_{s_{\tau}}(\rho)$, $\psi_{\rho,s_{\tau}}$, and $\psi_{\rho,s_{\tau}}'$, which corresponds to the evaluation of 3 two-dimensional integrals, and by optimizing over $\tau$. The normal approximation \eqref{eq:NA} can be obtained by evaluating $C(\rho)$ and $V(\rho)$, which corresponds to the evaluation of 2 two-dimensional integrals. The high-SNR normal approximation \cite[Th. 2]{LKD17} is available in closed form.

In Table~\ref{tb:complex_vs_time}, we summarize the computational complexities of the presented bounds and approximations. We further show the computational time (in seconds) required to numerically compute bounds and approximations on $R^{*}(L,T,\epsilon,\rho)$ for the parameters $L=30$, $T=10$, $\epsilon=10^{-6}$, and $\rho=6~\text{dB}$ on a PC-cluster node with 96 GB of RAM, powered by an Intel Xeon Gold 6130 processor. Observe that the computational complexity of the nonasymptotic bounds grows exponentially in~$L$, whereas the approximations proposed in this paper have a computational complexity that is independent of $L$. This is a significant reduction in computation cost, especially if $L$ is large. As a consequence, the saddlepoint approximations have a computational time that is about a factor of 20 (\rcus{} bound) or 17 (MC bound) smaller than that of the nonasymptotic bounds. The prefactor-and-error-exponent approximations have a similar complexity and computational time as the saddlepoint approximations, while the complexity and computational time of the error-exponent approximation is slightly smaller. The normal approximation has, by far, the lowest computational time, since no optimization over auxiliary parameters is required. Interestingly, the high-SNR normal approximation requires more computational time than the normal approximation, despite the fact that the former is available in closed form and the latter must be evaluated numerically. The reason is that the high-SNR approximation \eqref{eq:sec:first_first_moment_closed}  of capacity depends on a Gaussian hypergeometric function, whose evaluation is costly. Last but not least, it is worth pointing out that setting $s=1/(1+\tau)$ in the saddlepoint approximations and optimizing over $\tau$, which reduces the computational complexity to $5KN^2$, yields accurate results.
\subsection{Saddlepoint Approximations vs.\ Normal and Error-Exponent Approximations}\label{sec:NA_vs_saddle}
Intuitively, the error-exponent approximation \eqref{eq:err_approx} is accurate for small values of $\rho$ and $\epsilon$, whereas the normal approximation \eqref{eq:NA} is accurate for large values of $\rho$ and $\epsilon$. For example, we observe from Fig.~\ref{fig:rate_vs_Pe_0db} that, for $\rho=0$~dB, the normal approximation ``NA'' is accurate for $\epsilon>10^{-2}$. In contrast, for $\rho=0$~dB, the error-exponent approximation ``EEA" is accurate for error probabilities below $10^{-4}$. Similarly, we observe from Fig.~\ref{fig:rate_vs_Pe_6db} that, for $\rho=6$~dB, the normal approximation is accurate for error probabilities above $10^{-3}$, while the error-exponent approximation is accurate for $\epsilon<10^{-7}$. As noted in \cite{LKD17}, the high-SNR normal approximation is accurate for $\rho\geq 15$~dB, $L\geq 10$, and large values of $\epsilon$. In contrast, the saddlepoint approximations \eqref{eq:saddle_RCUs_numeric_gen} and \eqref{eq:MC_Saddle}, and the approximations \eqref{eq:rcus_SP_cor_2_th}, are accurate over the entire range of system parameters. Finally, the approximation \eqref{eq:err_approx_2} (``$L^{-\frac{1+\tau}{2}}$+EEA'') is accurate over the entire range of system parameters, albeit not as accurate as the saddlepoint approximations. 
That said, the normal approximations require a computational time that is two orders of magnitude lower than that of the remaining approximations. Furthermore, the high-SNR normal approximation is available in closed form, which makes it suitable for analytical studies.

In summary, the normal approximations are the go-to choice when the SNR and error probabilities are sufficiently large. For example, for an SNR of $6$~dB, this is the case for error probabilities above $10^{-3}$. Indeed, in this regime the normal approximations are reasonably accurate, and they can either be computed efficiently or are even available in closed form. In contrast, the saddlepoint approximations arise as easy-to-compute alternatives to the nonasymptotic bounds when one wishes to characterize the maximum coding rate for a large range of system parameters or for small SNR values and error probabilities, where the normal approximations are inaccurate. 
\section{Summary and Conclusions}\label{sec:conclusions}
In this paper, we applied the saddlepoint method to derive approximations of the MC and the \rcus{} bounds for single-antenna Rayleigh block-fading channels. While these approximations must still be evaluated numerically, they only require the evaluation of two-dimensional integrals. This is in contrast to the nonasymptotic MC and \rcus{} bounds, which require the evaluation of $(2L)$-dimensional integrals. Numerical evidence shows that the saddlepoint approximations are accurate for the entire range of system parameters for which the nonasymptotic bounds are computable. They thus arise as easy-to-compute alternatives to the nonasymptotic bounds.
\appendices
\section{Proofs of Proposition~\ref{pr:main_saddle} and Corollary~\ref{co:K_RCUs}}
\subsection{Proof of Proposition~\ref{pr:main_saddle}, Part~1}\label{app:proof_propMC}
The proof follows closely the steps by Feller \cite[Ch. XVI]{Feller}. Since we consider a random variable  $Z_{\ell,\theta}$ that depends on a parameter $\theta$, we provide here a self-contained proof proving uniformity in the extra parameter. Let $F_\theta$ denote the distribution of $Y_{\ell,\theta} \triangleq Z_{\ell,\theta}-\tilde{\gamma}$, where $\tilde{\gamma}\triangleq \gamma/n$. Then, the CGF of $Y_k$ is given by $\tilde{\psi}_\theta(\zeta) \triangleq \psi_\theta(\zeta)-\zeta\tilde{\gamma}$.
Consider the tilted random variable $\tiltRv$ with distribution
\begin{equation}\label{eq:tilting}
\tiltDist(x)=e^{-\tilde{\psi}_\theta(\tau)} \int_{-\infty}^{x}e^{\tau t} dF_\theta(t) = e^{-\psi_\theta(\tau)+\tau \tilde{\gamma}} \int_{-\infty}^{x}e^{\tau t} dF_\theta(t)
\end{equation}
where the parameter $\tau$ lies in $(-\zeta_0,\zeta_0)$.
Note that the exponential term $e^{-\psi_\theta(\tau)+\tau \tilde{\gamma}}$ on the right-hand side (RHS) of \eqref{eq:tilting} is a normalizing factor that guarantees that $\tiltDist$ is a distribution.

Let $v_{\theta,\tau}(\zeta)$ denote the MGF of the tilted random variable $\tiltRv$, which is given by
\begin{IEEEeqnarray}{lCl}
v_{\theta,\tau}(\zeta) &=& \int_{-\infty}^{\infty} e^{\zeta x} d\tiltDist(x) = \frac{m_\theta(\zeta+\tau)}{m_\theta(\tau)} e^{-\zeta\tilde{\gamma}}\label{eq:MGF_tilted}
\end{IEEEeqnarray}
where $m_\theta(\tau)$ is given in \eqref{eq:MGF_x_def}. Together with $\mathsf{E}\lrho{\tiltRv} = v_{\theta,\tau}'(0)$, this yields
\begin{IEEEeqnarray}{lCl}
\mathsf{E}\lrho{\tiltRv} &=& \psi_\theta'(\tau)-\tilde{\gamma}. \label{eq:first_mom_tilted_rv}
\end{IEEEeqnarray}
Let now $F_\theta^{\star n}$ denote the distribution of $\sum_{\ell=1}^n \lro{Z_{\ell,\theta}-\tilde{\gamma}}$, and let
 $\tiltDist^{\star n}$ denote the distribution of $\sum_{\ell=1}^n \tiltRv$. By \eqref{eq:tilting} and \eqref{eq:MGF_tilted}, the distributions $F_\theta^{\star n}$ and $\tiltDist^{\star n}$ again stand in the relationship \eqref{eq:tilting} except that the term $e^{-\psi_\theta(\tau)}$ is replaced by $e^{-n\psi_\theta(\tau)}$ and $\tilde{\gamma}$ is replaced by $n\tilde{\gamma}$. By inverting \eqref{eq:tilting}, we establish that
\begin{equation}\label{eq:prob_terms_Vstar}
\mathsf{P}\lrho{\sum\limits_{\ell=1}^n Z_{\ell,\theta} \geq \gamma} = e^{n\psi_\theta(\tau)-\tau\gamma} \int_0^\infty e^{-\tau y} d\tiltDist^{\star n}(y).
\end{equation}
Furthermore, by choosing $\tau$ such that $n\psi_{\theta}'(\tau) = \gamma$, it follows from \eqref{eq:first_mom_tilted_rv} that the distribution $\tiltDist^{\star n}$ has zero mean. We next substitute in \eqref{eq:prob_terms_Vstar} the distribution $\tiltDist^{\star n}$ by the zero-mean normal distribution with variance $n\psi_\theta''(\tau)$, denoted by $\mathfrak{N}_{n\psi_\theta''(\tau)}$, and analyze the error incurred by this substitution. To this end, we define
\begin{equation}\label{eq:A_tau}
A_\tau \triangleq e^{n\psi_\theta(\tau)-\tau\gamma} \int_0^\infty e^{-\tau y} d\mathfrak{N}_{n\psi_\theta''(\tau)}(y)
\end{equation}
which for $n\psi'(\tau)=\gamma$ can be evaluated as
\begin{IEEEeqnarray}{lCl}
	A_\tau &=& \frac{e^{n\lrho{\psi_\theta(\tau)-\tau\psi_\theta'(\tau)}}}{\sqrt{2\pi n\psi_\theta''(\tau)}}\int_{0}^{\infty} e^{-\tau y} e^{-\frac{y^2}{2n\psi_\theta''(\tau)}} dy  = e^{n\lrho{\psi_\theta(\tau)-\tau\psi_\theta'(\tau)+\frac{\tau^2}{2}\psi_\theta''(\tau)}} Q\lro{\tau\sqrt{n\psi_\theta''(\tau)}}.\IEEEyesnumber\label{eq:A_tau_expansion}
\end{IEEEeqnarray}

We continue by showing that the error incurred by substituting $\mathfrak{N}_{n\psi_\theta''(\tau)}$ for $\tiltDist^{\star n}$ in \eqref{eq:prob_terms_Vstar} is small. To do so, we note that integration by parts \cite[Ch. V.6, Eq. (6.1)]{Feller} yields
\begin{multline}
	\mathsf{P}\lrho{\sum\limits_{\ell=1}^n Z_\ell \geq n \psi_\theta'(\tau)}-A_\tau \\ 
	= e^{n[\psi_\theta(\tau)-\tau\psi_\theta'(\tau)]} \biggl[-\lro{\tiltDist^{\star n}(0)-\mathfrak{N}_{n\psi_\theta''(\tau)}(0)} +\tau \int_{0}^{\infty}\lro{\tiltDist^{\star n}(y)-\mathfrak{N}_{n\psi_\theta''(\tau)}(y)}e^{-\tau y}dy\biggr].\label{eq:step1_aprraising_diff}
\end{multline}
We next use \cite[Sec. XVI.4, Th.~1]{Feller} (stated as Lemma~\ref{lm:expansion_dist} below) to assess the error committed in \eqref{eq:step1_aprraising_diff}. 
To state Lemma~\ref{lm:expansion_dist}, we first introduce the following notation. Let $\tilde{Z}_{1,\theta},\dots,\tilde{Z}_{n,\theta}$ be a sequence of i.i.d., real-valued, zero-mean, random variables with one-dimensional probability distribution $\tilde{F}_\theta$ that depends on an extra parameter $\theta\in\Theta$. We denote the $k$-th moment for any possible value of $\theta\in\Theta$ by $\mu_{k,\theta}$ and we denote the second moment as $\mu_{2,\theta}=\sigma_\theta^2$.

For the distribution of the normalized $n$-fold convolution of a sequence of i.i.d., zero-mean, unit-variance random variables, we write
\begin{equation}\label{eq:F_unit_var}
\tilde{F}_{n,\theta}(x) = \tilde{F}_\theta^{\star n}(x\sigma_\theta\sqrt{n}).
\end{equation}
Note that $\tilde{F}_{n,\theta}$ has zero mean and unit variance. We denote by $\mathfrak{N}$ the zero-mean, unit-variance, normal distribution, and we denote by $\mathfrak{n}$ the zero-mean, unit-variance, normal pdf.
\begin{lemma}\label{lm:expansion_dist}
 Assume that the family of distributions $\tilde{F}_{n,\theta}$ (parametrized by $\theta$) is nonlattice. Further assume that
\begin{IEEEeqnarray}{rCl}
  \sup_{\theta\in\Theta} \mu_{4,\theta} &<& \infty\label{eq:moments_finite} \\
  \inf_{\theta\in\Theta} \sigma_{\theta} &>& 0. \label{eq:sec_moment_positive}
\end{IEEEeqnarray}
Then,  for every $\theta\in\Theta$ and $x\in\mathbb{R}$,
\begin{equation}\label{eq:expansions_distributions}
\tilde{F}_{n,\theta}(x) - \mathfrak{N}(x) = \frac{\mu_{3,\theta}}{6\sigma_\theta^3\sqrt{n}}(1-x^2)\mathfrak{n}(x) + o\lro{\frac{1}{\sqrt{n}}}
\end{equation}
where the $o(1/\sqrt{n})$ term is uniform in $x$ and $\theta$.
\end{lemma}
\begin{IEEEproof}
See Appendix~\ref{app:Proof_theorem_Feller}.
\end{IEEEproof}

We next use Lemma~\ref{lm:expansion_dist} to expand \eqref{eq:step1_aprraising_diff}. Recall that the family $F_\theta^{\star n}$ (parametrized by $\theta$) is nonlattice by assumption. Furthermore, as shown in Appendix~\ref{app:exp_tilt_lattice}, if a family of distributions is nonlattice, then so is the corresponding family of tilted distributions. Consequently, the family of distributions $\tiltDist^{\star n}$ (parametrized by $\theta$) is nonlattice, too. Note that the variable $y$ in \eqref{eq:step1_aprraising_diff} corresponds to $x\sigma_\theta\sqrt{n}$ in \eqref{eq:F_unit_var}. Hence, applying \eqref{eq:expansions_distributions} to \eqref{eq:step1_aprraising_diff} with $\tiltDist^{\star n}(y) = \tilde{F}_{n,\theta}(y/\sqrt{n\psi_{\theta}''(\tau)})$ and $\mathfrak{N}_{n\psi_{\theta}''(\tau)}(y) = \mathfrak{N} (y/\sqrt{n\psi_{\theta}''(\tau)})$, we obtain
\begin{IEEEeqnarray}{lCl}
\IEEEeqnarraymulticol{3}{l}{
	\mathsf{P}\lrho{\sum\limits_{\ell=1}^n Z_{\ell,\theta} \geq n \psi_\theta'(\tau)}-A_\tau}\IEEEnonumber*\\\,\,
	&=& e^{n[\psi_\theta(\tau)-\tau\psi_\theta'(\tau)]} \biggl[\frac{\psi_\theta'''(\tau)}{6\psi_\theta''(\tau)^{3/2}\sqrt{n}} \lro{-\frac{1}{\sqrt{2\pi}}+\frac{\tau^2n\psi_\theta''(\tau)}{\sqrt{2\pi}}-\tau^3\psi_\theta''(\tau)^{3/2}n^{3/2}\Qexp_\theta(\tau,\tau)} + o\lro{\frac{1}{\sqrt{n}}}\biggr]\\	\IEEEyesnumber\label{eq:difference_dist_end}
\end{IEEEeqnarray}
where we used that $\mathsf{Var}\lrho{\tiltRv}
= \psi_\theta''(\tau)$ and $\mathsf{E}\lrho{\lro{\tiltRv-\mathsf{E}\lrho{\tiltRv}}^3} = \psi_\theta'''(\tau)$. 

Substituting $A_\tau$ in \eqref{eq:A_tau_expansion} into \eqref{eq:difference_dist_end}, and recalling that $n\psi_{\theta}'(\tau) = \gamma$, we establish Part~1 of Proposition~\ref{pr:main_saddle}.
\subsection{Proof of Proposition~\ref{pr:main_saddle}, Part~2}\label{app:proof_propRCUs}
The proof of Part 2 follows along similar lines as the proof of Part 1. Hence, we will focus on describing what is different. Specifically, the left-hand-side (LHS) of \eqref{eq:saddlepoint_U} differs from the LHS of \eqref{eq:saddlepoint} by the additional term $\log U$. To account for this difference, we follow the same steps as Scarlett \emph{et al.} \cite[Appendix E]{Scarlett_2014}. Since in our setting the distribution of $Z_{\ell,\theta}$ depends on the parameter $\theta$, we repeat these steps in the following:
\begin{IEEEeqnarray}{lCl}\label{eq:prob_terms_Vstar_U}
\mathsf{P}\lrho{\sum\limits_{\ell=1}^n Z_{\ell,\theta} \geq \gamma + \log U} &=& e^{n\psi_\theta(\tau)-\tau\gamma} \int_0^1 \int_{\log u}^\infty e^{-\tau y} d\tiltDist^{\star n}(y) du\IEEEnonumber*\\
&=&e^{n\psi_\theta(\tau)-\tau\gamma}\lro{ \int_0^\infty e^{-\tau y} d\tiltDist^{\star n}(y) + \int_{-\infty}^0 e^{(1-\tau) y} d\tiltDist^{\star n}(y)}\IEEEyesnumber
\end{IEEEeqnarray}
where the second equality follows by changing the order of integration. We next proceed as in the proof of the previous part. The first term in \eqref{eq:prob_terms_Vstar_U} coincides with \eqref{eq:prob_terms_Vstar}. For the second term, we substitute the distribution $\tiltDist^{\star n}$ by the zero-mean normal distribution with variance $n\psi_\theta''(\tau)$ and analyze the error incurred by this substitution. To this end, we define
\begin{equation}\label{eq:A_tau_U}
\tilde{A}_\tau \triangleq e^{n\psi_\theta(\tau)-\tau\gamma} \int_{-\infty}^0 e^{(1-\tau) y} d\mathfrak{N}_{n\psi_\theta''(\tau)}(y)
\end{equation}
which for $n\psi'(\tau)=\gamma$ can be computed as
\begin{IEEEeqnarray}{lCl}
	\tilde{A}_\tau &=& e^{n\lrho{\psi_\theta(\tau)-\tau\psi_\theta'(\tau)+\frac{(1-\tau)^2}{2}\psi_\theta''(\tau)}} Q\lro{(1-\tau)\sqrt{n\psi_\theta''(\tau)}}.\IEEEyesnumber\label{eq:A_tau_expansion_U}
\end{IEEEeqnarray}

As we did in \eqref{eq:step1_aprraising_diff}, we next evaluate the error incurred by substituting $\tiltDist^{\star n}$ by $\mathfrak{N}_{n\psi_\theta''(\tau)}$. Indeed,
\begin{multline}
e^{n\psi_\theta(\tau)-\tau\gamma}\int_{-\infty}^0 e^{(1-\tau) y} d\tiltDist^{\star n}(y)-\tilde{A}_\tau = e^{n[\psi_\theta(\tau)-\tau\psi_\theta'(\tau)]} \biggl[ \frac{\psi_\theta'''(\tau)}{6\psi_\theta''(\tau)^{3/2}\sqrt{n}}\\ \times\lro{\frac{1}{\sqrt{2\pi}}-\frac{(1-\tau)^2n\psi_\theta''(\tau)}{\sqrt{2\pi}}+(1-\tau)^3\lro{n\psi_\theta''(\tau)}^{3/2}\Qexp_\theta(1-\tau,\tau)  } +o\lro{\frac{1}{\sqrt{n}}}\biggr]\label{eq:step1_aprraising_diff_U}
\end{multline}
which follows by integration by parts \cite[Ch. V.6, Eq. (6.1)]{Feller} and by Lemma~\ref{lm:expansion_dist}.

Combining \eqref{eq:prob_terms_Vstar_U} with \eqref{eq:prob_terms_Vstar}, \eqref{eq:saddlepoint}, \eqref{eq:A_tau_expansion_U}, and \eqref{eq:step1_aprraising_diff_U}, we obtain the desired result~\eqref{eq:saddlepoint_U}.
\subsection{Proof of Corollary~\ref{co:K_RCUs}}\label{app:proof_coRCUs}
Using \eqref{eq:prob_terms_Vstar_U} with \eqref{eq:A_tau} and \eqref{eq:A_tau_U}, and using a change of variable, we obtain
\begin{IEEEeqnarray}{lCl}
	\IEEEeqnarraymulticol{3}{l}{
	\mathsf{P}\lrho{\sum\limits_{\ell=1}^n Z_{\ell,\theta} \geq \gamma+\log U}-A_\tau-\tilde{A}_\tau}\IEEEnonumber*\\\,
	  &=& e^{n[\psi_\theta(\tau)-\tau\psi_\theta'(\tau)]} \biggl[\frac{1}{\sqrt{2\pi}}\frac{\psi_\theta'''(\tau)}{6\psi_\theta''(\tau)^{3/2}\sqrt{n}} \biggl(\int_{0}^{\infty}\tau\sqrt{n\psi_\theta''(\tau)}\lro{1-z^2} e^{-\tau\sqrt{n\psi_\theta''(\tau)}z -\frac{z^2}{2}} dz\\
	  && \qquad\qquad\qquad\quad+ \int_{0}^{\infty}\lro{1-\tau}\sqrt{n\psi_\theta''(\tau)}\lro{z^2-1} e^{-\lro{1-\tau}\sqrt{n\psi_\theta''(\tau)}z -\frac{z^2}{2}} dz\biggr) +o\lro{\frac{1}{\sqrt{n}}}\biggr].\IEEEeqnarraynumspace\IEEEyesnumber\label{eq:difference_dist_end_U_upper_2}
\end{IEEEeqnarray}
Keeping only the positive part of each integral, it follows that the RHS of \eqref{eq:difference_dist_end_U_upper_2} can be upper-bounded by
\begin{multline}
e^{n[\psi_\theta(\tau)-\tau\psi_\theta'(\tau)]} \biggl[\frac{1}{\sqrt{2\pi}}\frac{\psi_\theta'''(\tau)}{6\psi_\theta''(\tau)^{3/2}\sqrt{n}} \biggl(\int_{0}^{1}\tau\sqrt{n\psi_\theta''(\tau)}\lro{1-z^2} e^{-\tau\sqrt{n\psi_\theta''(\tau)}z -\frac{z^2}{2}} dz\\
{} +\int_{1}^{\infty}\lro{1-\tau}\sqrt{n\psi_\theta''(\tau)}\lro{z^2-1} e^{-\lro{1-\tau}\sqrt{n\psi_\theta''(\tau)}z -\frac{z^2}{2}} dz\biggr) +o\lro{\frac{1}{\sqrt{n}}}\biggr].\label{eq:difference_dist_end_U_upper_3}
\end{multline}
The first integral in \eqref{eq:difference_dist_end_U_upper_3} is upper-bounded by $1$. The second integral is upper-bounded by
\begin{IEEEeqnarray}{lCl}
 \frac{\lro{\lro{1-\tau}\sqrt{n\psi_\theta''(\tau)}\lro{\lro{1-\tau}\sqrt{n\psi_\theta''(\tau)}+2}+2}e^{-\lro{1-\tau}\sqrt{n\psi_\theta''(\tau)}}}{\lro{1-\tau}^2n\psi_\theta''(\tau)}.\IEEEyesnumber\label{eq:difference_dist_end_U_upper_4b}
\end{IEEEeqnarray}
If $\tau\in[0,\min\{\zeta_0,1-\delta\})$ for some arbitrary $\delta>0$ independent of $n$ and $\theta$, then the RHS of~\eqref{eq:difference_dist_end_U_upper_4b} vanishes faster than $1/\sqrt{n}$ uniformly in $\theta$. We thus obtain the upper bound
\begin{IEEEeqnarray}{lCl}
	\mathsf{P}\lrho{\sum\limits_{\ell=1}^n Z_\ell \geq n \psi_\theta'(\tau)+\log U}-A_\tau-\tilde{A}_\tau &\leq& e^{n[\psi_\theta(\tau)-\tau\psi_\theta'(\tau)]}\lrho{ \frac{1}{\sqrt{2\pi}}\frac{\psi_\theta'''(\tau)}{6\psi_\theta''(\tau)^{3/2}\sqrt{n}} + o\lro{\frac{1}{\sqrt{n}}}}\nonumber\\\IEEEyesnumber\label{eq:difference_dist_end_U_upper}
\end{IEEEeqnarray}
thereby proving Corollary~\ref{co:K_RCUs}.
\section{Proof of Lemma~\ref{lm:expansion_dist}}
\label{app:Proof_theorem_Feller}
The proof follows along similar lines as the proof of \cite[Ch. XVI.4, Th.~1]{Feller}. Notice that our result holds uniformly in the parameter $\theta$ of the distribution $\tilde{F}_\theta$, which makes the conditions of our lemma slightly more restrictive. Specifically, we require the fourth moment of $\tilde{F}_\theta$ to exist, whereas in the original theorem only the third moment is required to exist.

Let us denote the characteristic function of $\tilde{Z}_{\ell,\theta}\sim\tilde{F}_\theta$ by
\begin{equation}
	\tilde{\varphi}_\theta(\zeta) \triangleq \mathsf{E}\lrho{e^{\imath\zeta \tilde{Z}_{\ell,\theta}}}, \quad \zeta\in\mathbb{R}
\end{equation}
and define
\begin{equation}
G_\theta(x) \triangleq \mathfrak{N}(x)-\frac{\mu_{3,\theta}}{6\sigma_\theta^3\sqrt{n}}(x^2-1)\mathfrak{n}(x), \quad x\in\mathbb{R}.
\end{equation}
Note that \eqref{eq:moments_finite} implies that $\sup_{\theta\in\Theta}\lrvo{\mu_{3,\theta}} < \infty$ since, by Jensen's inequality, $\lrvo{\mu_{3,\theta}} \leq \mu_{4,\theta}^{3/4}$. Using this together with \eqref{eq:sec_moment_positive}, one can show that the first derivative of $G_\theta$ is bounded in $\theta\in\Theta$. Furthermore, the characteristic function of $G_\theta$ is
\begin{equation}\label{eq:Ch_function_G}
\gamma_\theta(\zeta) = e^{-\frac{1}{2}\zeta^2}\lrho{1+\frac{\mu_{3,\theta}}{6\sigma_\theta^3\sqrt{n}}(\imath\zeta)^3}.
\end{equation}
It follows that $G_\theta$ satisfies the conditions of \cite[Ch. XVI.3, Lemma~2]{Feller}, namely, that $\sup_{\theta\in\Theta,x\in\mathbb{R}}\bigl|G_\theta'(x)\bigr|\leq m$ for some positive constant $m$ and that $G_\theta$ has a continuously-differentiable characteristic function satisfying $\gamma_\theta(0) = 1$ and $\gamma_\theta'(0) = 0$. Then, the following inequality holds for all $x\in\mathbb{R}$ and $\Delta>0$ \cite[Ch. XVI.3, Eq. (3.13)]{Feller}:
\begin{equation}\label{eq:main_smoothing}
|\tilde{F}_{n,\theta}(x)-G(x)|\leq\frac{1}{\pi}\int_{-\Delta}^{\Delta}\Biggl|\frac{\varphi_\theta^n\lro{\frac{\zeta}{\sigma_\theta\sqrt{n}}}-\gamma_\theta(\zeta)}{\zeta}\Biggr|d\zeta + \frac{24m}{\pi \Delta}.
\end{equation}

Using \eqref{eq:main_smoothing} with $\Delta=a\sqrt{n}$, where the constant $a$ is chosen sufficiently large such that $\frac{24m}{\pi}<\varepsilon a$ for some arbitrary $\varepsilon>0$ independent of $x$ and $\theta$, we can write
\begin{equation}\label{eq:integral_proof_th_expans_dist}
|\tilde{F}_{n,\theta}(x)-G(x)|\leq \frac{1}{\pi}\int_{-a\sqrt{n}}^{a\sqrt{n}}\Biggl|\frac{\varphi_\theta^n\lro{\frac{\zeta}{\sigma_\theta\sqrt{n}}}-\gamma_\theta(\zeta)}{\zeta}\Biggr|d\zeta +\frac{\varepsilon}{\sqrt{n}}.
\end{equation}

It remains to show that the RHS of \eqref{eq:integral_proof_th_expans_dist} decays faster than $1/\sqrt{n}$ uniformly in $\theta$. To this end, we first note that, by assumption, the family of distributions $\tilde{F}_\theta$ (parametrized by $\theta$) is nonlattice, so $\sup_{\theta\in\Theta}|\varphi_\theta(\zeta)|$ is strictly smaller than $1$ for every $\zeta\neq 0$. Furthermore, as we shall argue below, \eqref{eq:moments_finite} implies that the function $\zeta\mapsto \sup_{\theta\in\Theta}\varphi_\theta(\zeta)$ is continuous. Consequently, there exists a number $q_{\delta,\bar{\zeta}}<1$ (independent of $\theta$) such that
\begin{equation}
\label{eq:qdeltazeta}
\sup_{\theta\in\Theta}|\varphi_\theta(\zeta)| \leq q_{\delta,\bar{\zeta}}, \quad \delta \leq |\zeta| \leq \bar{\zeta}
\end{equation}
for some arbitrary $\delta$ and $\bar{\zeta} \geq a / \inf_{\theta\in\Theta} \sigma_{\theta}$.

To prove that $\zeta\mapsto \sup_{\theta\in\Theta}\varphi_\theta(\zeta)$ is continuous, note that, by \cite[Ch.~XV.4, Lemma~2]{Feller},
\begin{equation}
\label{eq:cont_supvarphi}
\sup_{\theta\in\Theta} |\varphi_\theta'(\zeta)| \leq \sup_{\theta\in\Theta} \mathsf{E}\lrho{|\tilde{Z}_{\ell,\theta}|}, \quad \zeta\in\mathbb{R}.
\end{equation}
Moreover, for every $\zeta_1,\zeta_2\in\mathbb{R}$,
\begin{equation}
\left|\sup_{\theta\in\Theta} \varphi_{\theta}(\zeta_1) - \sup_{\theta\in\Theta}\varphi_{\theta}(\zeta_2)\right| \leq \sup_{\theta\in\Theta} \left|\varphi_{\theta}(\zeta_1)-\varphi_{\theta}(\zeta_2)\right| \leq \sup_{\theta\in\Theta} \mathsf{E}\lrho{|\tilde{Z}_{\theta}|}  |\zeta_1-\zeta_2| \label{eq:this_is_it}
\end{equation}
where the second inequality follows by the mean value theorem \cite[Th. 5.10]{rudin-principles}. 
Since the RHS of \eqref{eq:cont_supvarphi} is finite by \eqref{eq:moments_finite}, it follows that for every $\epsilon>0$ there exists a $\delta>0$ such that $|\zeta_1-\zeta_2| \leq \delta$ implies that the LHS of \eqref{eq:this_is_it} is bounded by $\epsilon$.
Thus, $\zeta\mapsto \sup_{\theta\in\Theta}\varphi_\theta(\zeta)$ is continuous.

We next bound the integral in \eqref{eq:integral_proof_th_expans_dist} by dividing the integration interval $|\zeta|\in(0,a\sqrt{n})$ into the two intervals $|\zeta|\in(\delta\sigma_{\theta}\sqrt{n},a\sqrt{n})$ and $|\zeta|\in(0,\delta\sigma_{\theta}\sqrt{n}]$ for some arbitrary $\delta>0$ that is independent of $x$ and $\theta$. The integral over the first interval can be bounded as
\begin{IEEEeqnarray}{lCl}
\IEEEeqnarraymulticol{3}{l}{\frac{1}{\pi}\int_{\delta\sigma_\theta\sqrt{n}<|\zeta|<a\sqrt{n}}\Biggl|\frac{\varphi_\theta^n\lro{\frac{\zeta}{\sigma_\theta\sqrt{n}}}-\gamma_\theta(\zeta)}{\zeta}\Biggr|d\zeta} \nonumber\\
& \leq & \frac{2}{\pi}\log\biggl(\frac{a}{\delta\sigma_{\theta}}\biggr)q_{\delta,\bar{\zeta}}^n + \frac{1}{\pi}\int_{\delta\sigma_\theta\sqrt{n}<|\zeta|<a\sqrt{n}} \biggl|\frac{\gamma_\theta(\zeta)}{\zeta}\biggr| d\zeta \nonumber\\
\qquad &\leq& \frac{2}{\pi}\log\biggl(\frac{a}{\delta \inf_{\theta\in\Theta}\sigma_{\theta}}\biggr) q_{\delta,\bar{\zeta}}^n + \frac{1}{\pi} \int_{|\zeta|>\delta\sqrt{n}\inf_{\theta\in\Theta}\sigma_\theta} \frac{e^{-\frac{1}{2}\zeta^2}}{|\zeta|}\lrho{1+\frac{\sup\limits_{\theta\in\Theta}\mu_{3,\theta}}{6\inf\limits_{\theta\in\Theta}\sigma_\theta^3\sqrt{n}}|\zeta|^3} d\zeta\IEEEeqnarraynumspace\label{eq:bound_contrib_big_values}
\end{IEEEeqnarray}
where the first inequality follows by upper-bounding the integrand using the triangle inequality and \eqref{eq:qdeltazeta}; the second inequality follows by upper-bounding $\gamma_\theta$, defined in \eqref{eq:Ch_function_G}, using the triangle inequality and by optimizing over $\theta\in\Theta$. The RHS of \eqref{eq:bound_contrib_big_values} tends to zero faster than any power of $1/n$ uniformly in $\theta$.

For the second interval, we express the integral as
\begin{equation}\label{eq:integral_proof_th_expans_dist_2}
\frac{1}{\pi}\int_{|\zeta|\leq\delta\sigma_\theta\sqrt{n}} e^{-\frac{1}{2}\zeta^2}\lrvo{\frac{\exp\lro{n\kappa_\theta\lro{\frac{\zeta}{\sigma_\theta\sqrt{n}}}}-1-\frac{n\mu_{3,\theta}}{6}\lro{\frac{\imath\zeta}{\sigma_\theta\sqrt{n}}}^3}{\zeta}} d\zeta
\end{equation}
where
\begin{equation}\label{eq:cum_plus_var_def}
	\kappa_\theta(\zeta) \triangleq \log\varphi_\theta(\zeta)+\frac{1}{2}\sigma_\theta^2\zeta^2.
\end{equation}
To bound \eqref{eq:integral_proof_th_expans_dist_2}, we will use that \cite[Ch. XVI.2, Eq. (2.8)]{Feller}
\begin{equation}\label{eq:bound_diff_exp}
\bigl|e^\alpha-1-\beta \bigr| \leq \biggl(|\alpha-\beta|+\frac{1}{2}\beta^2\biggr)e^\gamma, \quad \gamma\geq \max\lro{|\alpha|,|\beta|}.
\end{equation}
Recall that, by assumption \eqref{eq:moments_finite}, the fourth moment of $\tilde{F}_\theta$ is bounded. This implies that
\begin{equation}
\label{eq:third_Abs_moment}
\sup_{\theta\in\Theta}\int_{-\infty}^{\infty} |x|^{k} \mathrm{d} \tilde{F}_\theta(x)<\infty, \quad k=1,2,3
\end{equation}
since, by Jensen's inequality, $\mathsf{E}[|\tilde{Z}_{\ell,\theta}|^{k}] \leq \mu_{4,\theta}^{k/4}$, $k=1,2,3$.
Then, given an $\epsilon>0$ independent of $\theta$ and $\zeta$, it is possible to choose a $\delta$ (independent of $\theta$ and $\zeta$) such that, for $|\zeta|<\delta$,
\begin{equation}\label{eq:bound_third_order_expansion}
\Bigl|\kappa_\theta(\zeta)-\frac{1}{6}\mu_{3,\theta}(\imath\zeta)^3\Bigr|<\epsilon|\zeta|^3
\end{equation}
and
\begin{equation}\label{eq:bound_second_order_expansion}
|\kappa_\theta(\zeta)|<\frac{1}{4}\sigma_\theta^2\zeta^2, \qquad \Bigl|\frac{1}{6}\mu_{3,\theta}(\imath\zeta)^3\Bigr|\leq\frac{1}{4}\sigma_\theta^2\zeta^2.
\end{equation}
Indeed, after a Taylor series expansion of $\zeta\mapsto\kappa_\theta(\zeta)$ around $\zeta=0$, and noting that $\kappa_\theta(0)=\kappa_\theta'(0)=\kappa_\theta''(0)$, the LHS of \eqref{eq:bound_third_order_expansion} becomes
\begin{equation}\label{eq:term_third_to_bound}
\Bigl|\kappa_\theta(\zeta)-\frac{1}{6}\mu_{3,\theta}(\imath\zeta)^3\Bigr| = \Bigl|\frac{1}{6}\kappa_\theta'''(\tilde{\zeta})\zeta^3-\frac{1}{6}\mu_{3,\theta}(\imath\zeta)^3\Bigr|
\end{equation}
for some $\tilde{\zeta} \in (0,\zeta)$. Equation \eqref{eq:third_Abs_moment} implies that $\varphi_\theta'''(0)$ exists and \cite[Ch. XV.4, Lemma~2]{Feller}
\begin{equation}\label{eq:eqs_third}
\varphi_\theta'''(0) = \kappa_\theta'''(0) = \imath^3\mu_{3,\theta}.
\end{equation}
Furthermore, following the steps \eqref{eq:cont_supvarphi}--\eqref{eq:this_is_it}, it can be shown that, for every $\xi>0$,
\begin{equation}
\label{eq:taylor_phi}
 \sup_{\theta\in\Theta, |\zeta|<\xi}\Bigl|\varphi_{\theta}^{(k)}(\zeta)-\varphi_{\theta}^{(k)}(0)\Bigr| \leq \sup_{\theta\in\Theta} \mathsf{E}\lrho{|\tilde{Z}_{\ell,\theta}|^k} \xi, \quad k=0,1,2,3.
\end{equation}
By the definition of $\kappa_\theta$ in \eqref{eq:cum_plus_var_def}, the $k$-th derivative $\kappa_\theta^{(k)}$ is given by the ratio between a combination of derivatives of $\varphi_\theta$ up to order $k$ in the numerator, and $\varphi_\theta^k$ in the denominator. Since $\varphi_{\theta}(0)=1$, and since, by \eqref{eq:moments_finite}, $\mathsf{E}[|\tilde{Z}_{\ell,\theta}|^k]$ is bounded in $\theta$, it follows from~\eqref{eq:taylor_phi} that, for every $\epsilon$, there exists a $\delta>0$ satisfying $\sup_{\theta\in\Theta, |\zeta|<\delta}|\kappa_\theta'''(\zeta)-\kappa_\theta'''(0)|\leq 6\epsilon$.
Combining this with \eqref{eq:eqs_third}, we conclude that \eqref{eq:term_third_to_bound} can be bounded as
\begin{equation}\label{eq:term_third_bound_end}
 \Bigl|\frac{1}{6}\kappa_\theta'''(\tilde{\zeta})\zeta^3-\frac{1}{6}\mu_{3,\theta}(\imath\zeta)^3\Bigr| = \frac{1}{6}|\zeta|^3\Bigl|\kappa_\theta'''(\tilde{\zeta})-\imath^3\mu_{3,\theta} \Bigr| \leq \epsilon |\zeta|^3, \quad |\zeta|<\delta.
\end{equation}
 This proves \eqref{eq:bound_third_order_expansion}. The inequalities in \eqref{eq:bound_second_order_expansion} follow along similar lines.

Using \eqref{eq:bound_diff_exp} together with \eqref{eq:bound_third_order_expansion} and \eqref{eq:bound_second_order_expansion}, and replacing $\zeta$ by $\frac{\zeta}{n \sigma_\theta}$, we obtain that the integrand in \eqref{eq:integral_proof_th_expans_dist_2} is upper-bounded by
\begin{IEEEeqnarray}{lCl}
\frac{e^{-\frac{1}{4}\zeta^2}}{|\zeta|}\lro{\frac{\epsilon}{\sigma_\theta^3\sqrt{n}}|\zeta|^3+\frac{\mu_{3,\theta}^2}{72n}\zeta^6} &\leq& e^{-\frac{1}{4}\zeta^2}\lro{\frac{\epsilon}{\inf\limits_{\theta\in\Theta}\sigma_\theta^3\sqrt{n}}\zeta^2+\frac{\sup\limits_{\theta\in\Theta}\mu_{3,\theta}^2}{72n}|\zeta|^5}, \quad |\zeta|<\delta\sigma_\theta\sqrt{n}.\IEEEeqnarraynumspace
\end{IEEEeqnarray}
Integrating over $\zeta$, we conclude that \eqref{eq:integral_proof_th_expans_dist_2} decays faster than $1/\sqrt{n}$ uniformly in $x$ and $\theta$. Since the same is true for \eqref{eq:bound_contrib_big_values}, and since $\varepsilon$ is arbitrary, we obtain that the RHS of \eqref{eq:integral_proof_th_expans_dist} is $o(1/\sqrt{n})$ uniformly in $\theta$. Lemma~\ref{lm:expansion_dist} thus follows.
\section{Nonlattice Distributions and Exponential Tilting}\label{app:exp_tilt_lattice}
\begin{lemma}\label{lm:lattice}
	Let $\varphi_\theta$ denote the characteristic function of some distribution $F_{\theta}$, and let $\tilde{\varphi}_\theta$ denote the characteristic function of the tilted distribution $\vartheta_{\theta,\tau}$ (cf.\ \eqref{eq:tilting}). Then, for every $\zeta\neq 0$,
	\begin{equation}\label{eq:familiy_nonlattice}
	\sup\limits_{\theta\in\Theta}\lrvo{\varphi_\theta\lro{\zeta}}<1 \quad \Longrightarrow \quad \sup\limits_{\theta\in\Theta}\lrvo{\tilde{\varphi}_\theta(\zeta)} < 1.
	\end{equation}
	Thus, if a family of distributions is nonlattice, then so is the family of tilted distributions.
\end{lemma}
\begin{IEEEproof}
The characteristic function of the tilted random variable~$V_\theta\sim \vartheta_{\theta,\tau}$ can be written as
\begin{IEEEeqnarray}{lCl}
	\tilde{\varphi}_\theta(\zeta) &\triangleq& \int_{-\infty}^{\infty} e^{\imath\zeta x} d \vartheta_{\theta,\tau}(x) = \mathsf{E}\lrho{e^{(\imath\zeta + \tau)Z_\theta}} e^{-\imath\zeta\tilde{\gamma}} \frac{1}{m_\theta(\tau)}
\end{IEEEeqnarray}
where $m_\theta(\tau)$ denotes the MGF of $Z_\theta\sim F_\theta$. It follows that
\begin{IEEEeqnarray}{lCl}
	\lrvo{\tilde{\varphi}_\theta(\zeta)} &=&  \lrvo{\mathsf{E}\lrho{e^{(\imath\zeta + \tau)Z_\theta}}} \frac{1}{m_\theta(\tau)}.
\end{IEEEeqnarray}
We next note that there exists an $\alpha=e^{\imath\phi}$ such that
\begin{equation}
	\label{eq:abs_ch}
	\lrvo{\varphi_\theta(\zeta)} = \alpha \mathsf{E}\lrho{e^{\imath\zeta Z_\theta}} = \mathsf{E}\lrho{\cos\lro{\zeta Z_\theta + \phi}}.
\end{equation}
Likewise, there exists an $\tilde{\alpha}=e^{\imath\tilde{\phi}}$ such that	
\begin{equation}
	\lrvo{\tilde{\varphi}_\theta(\zeta)}  = \tilde{\alpha} \frac{\mathsf{E}\lrho{e^{(\imath\zeta + \tau)Z_\theta}}}{m_\theta(\tau)} = \frac{\mathsf{E}\lrho{e^{\tau Z_\theta}\cos\lro{\zeta Z_\theta + \tilde{\phi}}}}{m_\theta(\tau)}.
\end{equation}
Furthermore, we have $e^{\imath\tilde{\phi}}\mathsf{E}\lrho{e^{\imath\zeta Z_\theta}}=\mathsf{E}[\cos(\zeta Z_\theta + \tilde{\phi})]+\imath \mathsf{E}[\sin(\zeta Z_\theta + \tilde{\phi})]$, so
\begin{IEEEeqnarray}{lCl}
	\mathsf{E}\lrho{\cos\lro{\zeta Z_\theta + \tilde{\phi}}}^2
	&\leq& \lrvo{e^{\imath\tilde{\phi}}\mathsf{E}\lrho{e^{\imath\zeta Z_\theta}}}^2 
	= \mathsf{E}\lrho{\cos\lro{\zeta Z_\theta + \phi}}^2 \IEEEyesnumber
\end{IEEEeqnarray}
where the last step is due to \eqref{eq:abs_ch}. Since $\mathsf{E}\lrho{\cos\lro{\zeta Z_\theta + \phi}}$ is equal to $|\varphi_\theta(\zeta)|$ and hence nonnegative, it follows that
\begin{equation}\label{eq:phi_bound}
	\mathsf{E}\lrho{\cos\lro{\zeta Z_\theta + \tilde{\phi}}} \leq \mathsf{E}\lrho{\cos\lro{\zeta Z_\theta + \phi}}.
\end{equation}

Let $f(Z_\theta) \triangleq 1 - \cos\lro{\zeta Z_\theta+{\phi}}$ and $\tilde{f}(Z_\theta) \triangleq 1 - \cos\lro{\zeta Z_\theta+{\tilde{\phi}}}$. The LHS of \eqref{eq:familiy_nonlattice} is equivalent to $\inf_{\theta\in\Theta}\mathsf{E}\lrho{f(Z_\theta)}>0$.
Similarly, the RHS of \eqref{eq:familiy_nonlattice} is implied by $\inf_{\theta\in\Theta}\mathsf{E}[e^{\tau Z_\theta}\tilde{f}(Z_\theta)]>0$
because
\begin{IEEEeqnarray}{lCl}
	1- \sup_{\theta\in\Theta}\frac{\mathsf{E}\lrho{e^{\tau Z_\theta}\cos\lro{\zeta Z_\theta + \tilde{\phi}}}}{m_\theta(\tau)}
	&\geq&  \frac{\inf_{\theta\in\Theta}\mathsf{E}\lrho{e^{\tau Z_\theta}\tilde{f}(Z_\theta)}}{\sup_{\theta\in\Theta}m_\theta(\tau)} \IEEEyesnumber\label{eq:phi_tilted_max}
\end{IEEEeqnarray}
and $m_\theta$ is bounded by assumption \eqref{eq:MGF_unif_bounded}.

We next show that
\begin{equation}
	\label{eq:tilted_lattice_f}
	\inf_{\theta\in\Theta}\mathsf{E}\lrho{e^{\tau Z_\theta}\tilde{f}(Z_\theta)}=0 \quad \Longrightarrow \quad \inf_{\theta\in\Theta}\mathsf{E}\lrho{\tilde{f}(Z_\theta)} =0.
\end{equation}
We then note that, by \eqref{eq:phi_bound}, $\mathsf{E}[\tilde{f}(Z_\theta)]  \geq \mathsf{E}\lrho{{f}(Z_\theta)}$. Since $f(\cdot)$ is nonnegative, \mbox{$\inf_{\theta\in\Theta}\mathsf{E}[\tilde{f}(Z_\theta)] =0$}
thus implies that $\inf_{\theta\in\Theta}\mathsf{E}[{f}(Z_\theta)]=0$. Hence, by reverse logic,
\begin{equation}
	\inf_{\theta\in\Theta}\mathsf{E}\lrho{{f}(Z_\theta)}>0 \quad \Longrightarrow \quad\inf_{\theta\in\Theta}\mathsf{E}\lrho{e^{\tau Z_\theta}\tilde{f}(Z_\theta)}>0
\end{equation}
which concludes the proof of Lemma~\ref{lm:lattice}.

To prove \eqref{eq:tilted_lattice_f}, we first note that, for every arbitrary $\delta>0$,
\begin{IEEEeqnarray}{lCl}
	\mathsf{E}\lrho{e^{\tau Z_\theta}\tilde{f}(Z_\theta)}
	&\geq& \mathsf{E}\lrho{e^{\tau Z_\theta}\tilde{f}(Z_\theta)\mathbb{I}\lrbo{\lrvo{Z_\theta}\leq \delta}}\geq \mathsf{E}\lrho{\tilde{f}(Z_\theta)\mathbb{I}\lrbo{\lrvo{Z_\theta}\leq \delta}}e^{-\tau \delta}\IEEEyesnumber\label{eq:fXeX}
\end{IEEEeqnarray}
where $\mathbb{I}\{\cdot\}$ denotes the indicator function. Furthermore,
\begin{IEEEeqnarray}{lCl}
	\mathsf{E}\lrho{\tilde{f}(Z_\theta)} & = & \mathsf{E}\lrho{\tilde{f}(Z_\theta)\mathbb{I}\lrbo{\lrvo{Z_\theta}\leq \delta}} + \mathsf{E}\lrho{\tilde{f}(Z_\theta)\mathbb{I}\lrbo{\lrvo{Z_\theta}> \delta}} \nonumber\\
	& \leq & \mathsf{E}\lrho{\tilde{f}(Z_\theta)\mathbb{I}\lrbo{\lrvo{Z_\theta}\leq \delta}} +  2\frac{\sup_{\theta\in\Theta}\mathsf{E}\lrho{Z_\theta^2}}{\delta^2} \label{eq:fX}
\end{IEEEeqnarray}
where the inequality follows because $\tilde{f}(Z_\theta)$ is upper-bounded by $2$ and from Chebyshev's inequality. Combining \eqref{eq:fXeX} with \eqref{eq:fX}, and using that $\tilde{f}(\cdot)$ is nonnegative, we thus obtain that
\begin{IEEEeqnarray}{lCl}
	0 \leq \inf_{\theta\in\Theta}\mathsf{E}\lrho{\tilde{f}(Z_\theta)} &\leq& 	e^{\tau\delta}\inf_{\theta\in\Theta}\mathsf{E}\lrho{e^{\tau Z_\theta}\tilde{f}(Z_\theta)} +  2\frac{\sup_{\theta\in\Theta}\mathsf{E}\lrho{Z_\theta^2}}{\delta^2} .\IEEEyesnumber\label{eq:fXeX_sofX}
\end{IEEEeqnarray}
By assumption~\eqref{eq:MGF_unif_bounded}, we have that $\sup_{\theta\in\Theta}\mathsf{E}\lrho{Z_\theta^2}<\infty$. Consequently, if $\inf_{\theta\in\Theta}\mathsf{E}[e^{\tau Z_\theta}\tilde{f}(Z_\theta)]=0$ then we obtain that $\inf_{\theta\in\Theta}\mathsf{E}[\tilde{f}(Z_\theta)] = 0$ upon letting $\delta\to\infty$. This yields \eqref{eq:tilted_lattice_f}.
\end{IEEEproof}
\section{The Second Derivative of the CGF Is Bounded Away from Zero}\label{app:CGF2_boundedaway}
\begin{lemma}\label{lm:CGF2_bounded_away}
Assume that $Z_\theta$ is zero mean and its MGF and CGF satisfy \eqref{eq:MGF_unif_bounded} and
\begin{IEEEeqnarray}{lCl}\label{eq:assumption_var_boundedaway}
	\inf_{\theta\in\Theta} \psi_\theta''(0) > 0.
\end{IEEEeqnarray}
Then, for every $\zeta_0>0$,
\begin{equation}\label{eq:second_CGF_boundedaway}
	\inf_{\theta\in\Theta, |\zeta|<\zeta_0} \psi_{\theta}''(\zeta) >0.
\end{equation}
\end{lemma}
\begin{IEEEproof}
The second derivative of the CGF is given by
\begin{IEEEeqnarray}{lCl}
	\psi_{\theta}''(\zeta)		& = &  \frac{\mathsf{E}\lrho{Z_\theta^2 e^{\zeta Z_\theta}}\mathsf{E}\lrho{e^{\zeta Z_\theta}}- \mathsf{E}\lrho{Z_\theta e^{\zeta Z_\theta}}^2}{m_\theta(\zeta)^2}. \IEEEeqnarraynumspace\IEEEyesnumber\label{eq:second_CGF_rho_s}
\end{IEEEeqnarray}
By~\eqref{eq:MGF_unif_bounded}, the denominator on the RHS of \eqref{eq:second_CGF_rho_s} is bounded. Thus, to obtain \eqref{eq:second_CGF_boundedaway}, it suffices to show that the numerator on the RHS of \eqref{eq:second_CGF_rho_s} is bounded away from zero.	To shorten notation, we next define $A \triangleq Z_\theta e^{\frac{\zeta}{2}Z_\theta}$, $B \triangleq e^{\frac{\zeta}{2}Z_\theta}$, $\sigma_A^2\triangleq \mathsf{E}\lrho{A^2}$, $\sigma_B^2\triangleq \mathsf{E}\lrho{B^2}$, $\overline{\lambda}_{A,B} \triangleq \frac{A}{\sigma_A}+\frac{B}{\sigma_B}$ and $\underline{\lambda}_{A,B} \triangleq \frac{A}{\sigma_A}-\frac{B}{\sigma_B}$.
Then, the numerator on the RHS of \eqref{eq:second_CGF_rho_s} becomes $\sigma_A^2 \sigma_B^2- \mathsf{E}\lrho{AB}^2$. By following the proof of the Cauchy-Schwarz inequality \cite[Th. 3.3.1]{Lapidoth09}, it can be shown that
\begin{equation}
	\lrvo{\mathsf{E}\lrho{A B}} \leq \sigma_A \sigma_B \mathsf{K}\IEEEyesnumber\label{eq:abs_EAB}
\end{equation}
	where
\begin{equation}
	\mathsf{K} \triangleq \max\left\{\lro{1-\frac{1}{2}\inf_{ \theta\in\Theta, |\zeta|<\zeta_0}\mathsf{E}\lrho{\underline{\lambda}_{A,B}^2}}^+,\lro{1-\frac{1}{2}\inf_{\theta\in\Theta, |\zeta|<\zeta_0}\mathsf{E}\lrho{\overline{\lambda}_{A,B}^2}}^+\right\}
\end{equation}
and $(x)^+\triangleq \max\{0,x\}$. Using \eqref{eq:abs_EAB}, we can lower-bound the numerator on the RHS of \eqref{eq:second_CGF_rho_s}~as
\begin{equation}
	\sigma_A^2 \sigma_B^2- \mathsf{E}\lrho{AB}^2
	\geq (1-\mathsf{K}^2) \inf_{\theta\in\Theta, |\zeta|<\zeta_0}\sigma_A^2 \inf_{\theta\in\Theta, |\zeta|<\zeta_0} \sigma_B^2  \geq (1-\mathsf{K}^2) \inf_{ \theta\in\Theta,|\zeta|<\zeta_0}\sigma_A^2 \label{eq:inf_second_CGF_AB_L}
\end{equation}
where the second inequality follows because $Z_{\theta}$ is zero mean by assumption, so Jensen's inequality gives $\sigma_B^2= \mathsf{E}\lrho{e^{\zeta Z_\theta}}\geq 1$. Thus, in order to show that the numerator on the RHS of \eqref{eq:second_CGF_rho_s} is bounded away from zero, it remains to prove that
\begin{IEEEeqnarray}{rCl}
	\inf_{\theta\in\Theta,|\zeta|<\zeta_0} \sigma_A^2 &>&0\IEEEyesnumber\IEEEyessubnumber*\label{eq:LA2}\\
	\inf_{ \theta\in\Theta, |\zeta|<\zeta_0} \mathsf{E}\lrho{\underline{\lambda}_{A,B}^2} &>&0 \label{eq:LdiffAB}\\
	\inf_{ \theta\in\Theta, |\zeta|<\zeta_0 }\mathsf{E}\lrho{\overline{\lambda}_{A,B}^2} &>&0. \label{eq:LsumAB}
\end{IEEEeqnarray}

To prove \eqref{eq:LA2}, recall that $\sigma_A^2=\mathsf{E}[Z_\theta^2 e^{\zeta Z_\theta}]$. We next show that
\begin{equation}
	\inf_{\theta\in\Theta, |\zeta|<\zeta_0}\mathsf{E}\lrho{Z_\theta^2 e^{\zeta Z_\theta}}=0 \quad \Longrightarrow \quad \inf_{\theta\in\Theta}\mathsf{E}\lrho{Z_\theta^2}=0. \label{eq:reverse_logic_A}
\end{equation}
Since $\mathsf{E}\lrho{Z_\theta^2}=\psi_{\theta}''(0)$, it follows by assumption \eqref{eq:assumption_var_boundedaway} that the RHS of \eqref{eq:reverse_logic_A} cannot be true. Hence, by reverse logic, $\inf_{\theta\in\Theta, |\zeta|<\zeta_0}\mathsf{E}[Z_\theta^2 e^{\zeta Z_\theta}]>0$, which is \eqref{eq:LA2}.

To prove \eqref{eq:reverse_logic_A}, we follow along the lines of \eqref{eq:fXeX}--\eqref{eq:fXeX_sofX}. Indeed, for every arbitrary $\delta>0$,
\begin{IEEEeqnarray}{lCl}
	\mathsf{E}\lrho{Z_\theta^2 e^{\zeta Z_\theta}}
	&\geq& \mathsf{E}\lrho{Z_\theta^2\mathbb{I}\lrbo{\lrvo{Z_\theta}\leq \delta}}e^{-\zeta \delta}.\IEEEyesnumber\label{eq:X2eX}
\end{IEEEeqnarray}
Furthermore,
\begin{IEEEeqnarray}{lCl}
	\mathsf{E}\lrho{Z_\theta^2} 
	&\leq& \mathsf{E}\lrho{Z_\theta^2\mathbb{I}\lrbo{\lrvo{Z_\theta}\leq \delta}} +  \frac{\sqrt{\sup_{\theta\in\Theta}\mathsf{E}\lrho{Z_\theta^4} \sup_{\theta\in\Theta}\mathsf{E}\lrho{Z_\theta^2}}}{\delta}\IEEEyesnumber\label{eq:X2}
\end{IEEEeqnarray}
by the Cauchy-Schwarz inequality and Chebyshev's inequality. Combining \eqref{eq:X2eX} with \eqref{eq:X2}, and using that $Z_\theta^2$ is nonnegative, it follows that
\begin{IEEEeqnarray}{lCl}
	0 \leq \inf_{\theta\in\Theta}\mathsf{E}\lrho{Z_\theta^2}	&\leq& 	e^{\zeta_0\delta}\inf_{\theta\in\Theta, |\zeta|<\zeta_0}\mathsf{E}\lrho{Z_\theta^2 e^{\zeta Z_\theta}} +   \frac{\sqrt{\sup_{\theta\in\Theta}\mathsf{E}\lrho{Z_\theta^4} \sup_{\theta\in\Theta}\mathsf{E}\lrho{Z_\theta^2}}}{\delta} .\IEEEyesnumber\label{eq:X2eX_soX2}
\end{IEEEeqnarray}
By assumption~\eqref{eq:MGF_unif_bounded}, we have that $\sup_{\theta\in\Theta}\mathsf{E}\lrho{Z_\theta^k}<\infty$ for $k=2,4$. Consequently, if $\inf_{\theta\in\Theta, |\zeta|<\zeta_0}\mathsf{E}\lrho{Z_\theta^2 e^{\zeta Z_\theta}}=0$ then we obtain that \mbox{$\inf_{\theta\in\Theta}\mathsf{E}\lrho{Z_\theta^2}=0$} upon letting $\delta\to\infty$. This yields \eqref{eq:reverse_logic_A}.

We next prove \eqref{eq:LdiffAB}. To do so, we write $\mathsf{E}[\underline{\lambda}_{A,B}^2]$ as $\mathsf{E}[(A-B \eta_{\theta,\zeta})^2]/\sigma_A^2$, where $\eta_{\theta,\zeta}\triangleq \sigma_A/\sigma_B$. It can be shown that $\sigma^2_A=m''_{\theta}(\zeta)$. Consequently, $\sigma_A$ is bounded by assumption~\eqref{eq:MGF_unif_bounded}. To prove \eqref{eq:LdiffAB}, it thus remains to show that
\begin{equation}
	\inf_{ \theta\in\Theta, |\zeta|<\zeta_0}\mathsf{E}\lrho{
	\lro{A-B\eta_{\theta,\zeta}}^2} = \inf_{\theta\in\Theta, |\zeta|<\zeta_0} \mathsf{E}\lrho{e^{\zeta Z_\theta}\lro{Z_\theta-\eta_{\theta,\zeta}}^2} >0.\label{eq:inf_Exp_diff}
\end{equation}
To this end, we first note that the steps \eqref{eq:X2eX}--\eqref{eq:X2eX_soX2} with $Z_\theta^2$ replaced by $\lro{Z_\theta-\eta_{\theta,\zeta}}^2$ yield
\begin{equation}
	\label{eq:reverse_logic_B}
	\inf_{\theta\in\Theta, |\zeta|<\zeta_0}\mathsf{E}\lrho{e^{\zeta Z_\theta}\lro{Z_\theta - \eta_{\theta,\zeta}}^2}=0 \quad \Longrightarrow \quad \inf_{\theta\in\Theta, |\zeta|<\zeta_0}\mathsf{E}\lrho{\lro{Z_\theta - \eta_{\theta,\zeta}}^2}=0.
\end{equation}
Since $Z_\theta$ is zero mean, we further have that $\mathsf{E}[\lro{Z_\theta - \eta_{\theta,\zeta}}^2]\geq\mathsf{E}\lrho{Z_\theta^2}$, so if $\mathsf{E}[(Z_\theta - \eta_{\theta,\zeta})^2]$ is zero, then so is $\mathsf{E}\lrho{Z_\theta^2}$. Since, by assumption \eqref{eq:assumption_var_boundedaway}, we have $\inf_{\theta\in\Theta}\mathsf{E}\lrho{Z_\theta^2}>0$, the inequality in \eqref{eq:inf_Exp_diff} follows from \eqref{eq:reverse_logic_B} by reverse logic.

Finally, \eqref{eq:LsumAB} follows from the same steps as the ones used to show \eqref{eq:LdiffAB}, but with $\eta_{\theta,\zeta}$ replaced by $-\eta_{\theta,\zeta}$.
\end{IEEEproof}

\bibliographystyle{IEEEtran}

\end{document}